\documentclass[8pt,preprint]{aastex}
\usepackage[]{aastexug,natbib,epsfig,graphicx,rotating,lscape,rotfloat,subfigure,lineno,textcomp,longtable}
\def\arcsec{$^{\prime\prime}$}
\def\arcmin{$^{\prime}$}
\def\degree{$^{\circ}$}
\newcommand{\NtwoH}{N$_{2}$H$^{+}$}
\newcommand{\HCO}{HCO$^{+}$}
\newcommand{\vlsr}{$V_{\mathrm{lsr}}$}

\newcommand{\Msun}{\mbox{${M}_{\sun}$}}
\newcommand{\Lsun}{\mbox{${L}_{\sun}$}}
\newcommand{\Jybm}{Jy~beam$^{-1}$}

\newcommand{\Jybmkms}{Jy~beam$^{-1}$~km~s$^{-1}$}
\newcommand{\kms}{km~s$^{-1}$}
\newcommand{\kmsppc}{km~s$^{-1}$~pc$^{-1}$}
\newcommand{\Av}{$A_{V}$}

\begin{document}

\title{CARMA LARGE AREA STAR FORMATION SURVEY: DENSE GAS IN THE YOUNG
  L1451 REGION OF PERSEUS}

\shorttitle{CARMA Large Area Star Formation Survey: L1451}

\tiny
\author{
Shaye Storm\altaffilmark{1,2}, 
Lee G. Mundy\altaffilmark{2},
Katherine I. Lee\altaffilmark{1,2},
Manuel Fern\'{a}ndez-L\'{o}pez\altaffilmark{3,4},
Leslie W. Looney\altaffilmark{3},
Peter Teuben\altaffilmark{2},
H\'{e}ctor G. Arce\altaffilmark{5},
Erik W. Rosolowsky\altaffilmark{6},
Aaron M. Meisner \altaffilmark{7},
Andrea Isella\altaffilmark{8},
Jens Kauffmann\altaffilmark{9},
Yancy L. Shirley\altaffilmark{10},
Woojin Kwon\altaffilmark{11},
Adele L. Plunkett\altaffilmark{12},
Marc W. Pound\altaffilmark{2},
Dominique M. Segura-Cox\altaffilmark{3},
Konstantinos Tassis\altaffilmark{13,14},
John J. Tobin\altaffilmark{15},
Nikolaus H. Volgenau\altaffilmark{16},
Richard M. Crutcher\altaffilmark{3},
Leonardo Testi\altaffilmark{17}
}
\altaffiltext{1}{\tiny Harvard-Smithsonian Center for Astrophysics, 60
  Garden Street, Cambridge, MA 02138, USA}
\altaffiltext{2}{\tiny Department of Astronomy, University of Maryland,
  College Park, MD 20742, USA; sstorm@astro.umd.edu}
\altaffiltext{3}{\tiny Department of Astronomy, University of Illinois at
  Urbana--Champaign, 1002 West Green Street, Urbana, IL 61801, USA}
\altaffiltext{4}{\tiny Instituto Argentino de Radioastronom{\'{\i}}a, CCT-La
  Plata (CONICET), C.C.5, 1894, Villa Elisa, Argentina}
\altaffiltext{5}{\tiny Department of Astronomy, Yale University, P.O.~Box
  208101, New Haven, CT 06520-8101, USA}
\altaffiltext{6}{\tiny University of Alberta, Department of Physics, 4-181
  CCIS, Edmonton AB T6G 2E1, Canada}
\altaffiltext{7}{\tiny Lawrence Berkeley National Laboratory and Berkeley
  Center for Cosmological Physics, Berkeley, CA 94720, USA}
\altaffiltext{8}{\tiny Physics \& Astronomy Department, Rice University,
  P.O. Box 1892, Houston, TX 77251-1892, USA}
\altaffiltext{9}{\tiny Max Planck Institut f\"{u}r Radioastronomie, Auf dem
  H\"{u}gel 69 D–53121, Bonn Germany}
\altaffiltext{10}{\tiny Steward Observatory, 933 North Cherry Avenue, Tucson,
  AZ 85721, USA}
\altaffiltext{11}{\tiny Korea Astronomy and Space Science Institute,776
  Daedeok-daero, Yuseong-gu, Daejeon 305-348, Republic of Korea}
\altaffiltext{12}{\tiny European Southern Observatory, Av. Alonso de Cordova
  3107, Vitacura, Santiago de Chile}
\altaffiltext{13}{\tiny Department of Physics and Institute of Theoretical
  \& Computational Physics, University of Crete, PO Box 2208, GR-710
  03, Heraklion, Crete, Greece}
\altaffiltext{14}{\tiny Foundation for Research and Technology - Hellas,
  IESL, Voutes, 7110 Heraklion, Greece}
\altaffiltext{15}{\tiny Leiden Observatory, 540 J.H. Oort Building, Niels
  Bohrweg 2, NL-2333 CA Leiden, The Netherlands}
\altaffiltext{16}{\tiny Las Cumbres Observatory Global Telescope Network,
  Inc. 6740 Cortona Drive, Suite 102 Goleta, CA 93117, USA}
\altaffiltext{17}{\tiny ESO, Karl-Schwarzschild-Strasse 2 D-85748 Garching
  bei M\"{u}nchen, Germany}

\email{Accepted to The Astrophysical Journal (ApJ); June 25, 2016}
\vspace{-0.2cm}
\email{(see published version for full-resolution figures)}

\normalsize
\begin{abstract}
 We present a 3~mm spectral line and continuum survey of L1451 in the
 Perseus Molecular Cloud. These observations are from the CARMA Large
 Area Star Formation Survey (CLASSy), which also imaged Barnard~1,
 NGC~1333, Serpens Main and Serpens South. L1451 is the survey region
 with the lowest level of star formation activity---it contains no
 confirmed protostars. \HCO{}, HCN, and \NtwoH{} ($J=1\rightarrow0$)
 are all detected throughout the region, with \HCO{} the most
 spatially widespread, and molecular emission seen toward 90\% of the
 area above N(H$_{2}$) column densities of
 1.9$\times$10$^{21}$~cm$^{-2}$.  \HCO{} has the broadest velocity
 dispersion, near 0.3~\kms{} on average, compared to $\sim$0.15~\kms{}
 for the other molecules, thus representing a range from supersonic to
 subsonic gas motions.  Our non-binary dendrogram analysis reveals
 that the dense gas traced by each molecule has similar hierarchical
 structure, and that gas surrounding the candidate first hydrostatic
 core (FHSC), L1451-mm, and other previously detected single-dish
 continuum clumps have similar hierarchical structure; this suggests
 that different sub-regions of L1451 are fragmenting on the pathway to
 forming young stars.  We determined the three-dimensional morphology
 of the largest detectable dense gas structures to be relatively
 ellipsoidal compared to other CLASSy regions, which appeared more
 flattened at largest scales. A virial analysis shows the most
 centrally condensed dust structures are likely unstable against
 collapse. Additionally, we identify a new spherical, centrally
 condensed \NtwoH{} feature that could be a new FHSC candidate. The
 overall results suggest L1451 is a young region starting to form its
 generation of stars within turbulent, hierarchical structures.

\end{abstract}

\section{Introduction}
\label{sec:ch3sec1}
The star formation process in a molecular cloud starts well before
protostars are detectable at infrared wavelengths. In general, it
begins with the formation of the molecular cloud that may span tens of
parsecs \citep{1999ARA&A..37..311E, 2004ARA&A..42..211E,
  2007ARAA..45..565M}; it continues as structure and density
enhancements are created by the interaction of turbulence, gravity,
and magnetic fields at parsec scales \citep{2007ARAA..45..565M,
  2012ARA&A..50...29C}, and it progresses until prestellar core
collapse occurs at 0.01--0.1~pc scales \citep{2007prpl.conf...17D,
  2007ARA&A..45..339B}. Once a first generation of protostars is
formed within those dense cores, the young stars can feed energy back
into the cloud and impact subsequent star formation that may occur
\citep{2007ApJ...662..395N, 2009ApJ...695.1376C, 2014ApJ...783..115N}.
A full understanding of how turbulence, gravity, and magnetic fields
control the star formation process requires observations that span
cloud to core spatial scales at these distinct evolutionary stages.

An individual molecular cloud can be a great testbed for studying the
star formation process across space and time if it is sufficiently
close to get better than 0.1~pc resolution, and if it contains regions
with distinct evolutionary stages. The Perseus Molecular Cloud is a
nearby example of such a cloud. The regions of Perseus with infrared
detections of young stellar objects (YSOs) span a range of
evolutionary epochs based on YSO statistics from the c2d Legacy
project \citep{2008ApJ...683..822J,2009ApJS..181..321E}. For example,
the IC~348 region has 121 YSOs, with 9.1\% being Class I or younger;
the NGC~1333 region has 102 YSOs, with 34\% being Class I or younger;
Barnard~1 region has 9 YSOs, with 89\% being Class I or younger.
Regions without current protostellar activity also exist within
Perseus. The B1-E region may be forming a first generation of dense
cores \citep{2012AA...540A..10S}, and the L1451 region has a single
detection of a compact continuum core, which is a candidate first
hydrostatic core (FHSC) \citep{2011ApJ...743..201P}.

The CARMA Large Area Star Formation Survey (CLASSy) observed the dense
gas in three evolutionary distinct regions within Perseus
\citep{2014ApJ...794..165S} and two regions within Serpens
\citep{2014ApJ...797...76L,2014ApJ...790L..19F} with high angular and
velocity resolution. The observations enable a high resolution study
of the structure and kinematics of star forming material at different
epochs. From early to late stages of evolution (based on the ratio of
Class II and older to Class I and younger YSOs), the Perseus regions
of CLASSy are L1451, Barnard~1, and NGC~1333. The youngest region,
L1451, probes cloud conditions during the origin of clumps and stars;
the more evolved Barnard~1 region probes cloud conditions when a
relatively small number of protostars are formed, and the active
NGC~1333 region probes cloud conditions when dozens of clustered
protostars are driving outflows back into the cloud.  Details of
CLASSy, along with an analysis of Barnard~1, can be found in
\citet{2014ApJ...794..165S} (referred to as Paper~I in the sections
below).

This paper focuses on the L1451 region. L1451 is located $\sim$5.5~pc
to the southwest of NGC~1333 (see Figure~1 of Paper~I).  The region
has been surveyed at a number of wavelengths as reported in the
literature and summarized below. It contains no
\textit{Spitzer}-identified YSOs at IRAC or MIPS wavelengths
\citep{2008ApJ...683..822J}\footnote{The \textit{Spitzer} c2d YSO
  sample is 90\% complete down to 0.05~\Lsun{} for clouds at 260~pc
  \citep{2009ApJS..181..321E}.}. \citet{2005A&A...440..151H} and
\citet{2006ApJ...646.1009K} did not identify any cores in their JCMT
SCUBA 850~$\mu$m survey. There are four 1.1~mm sources identified in
the Bolocam Survey \citep{2006ApJ...638..293E} that are classified as
``starless'' cores in \citet{2008ApJ...684.1240E}: PerBolo 1, 2, 4,
and 6. \citet{2011ApJ...743..201P} used 3~mm CARMA and 1.3~mm SMA
observations to show that PerBolo~2 is not a starless core, but likely
a core with an embedded YSO or a FHSC.

The Bolocam cores within L1451 are colder and less dense than the
average Bolocam cores within Perseus.  The visual extinction (\Av) of
the four Bolocam sources ranges from 8 to 11 magnitudes, while the
mean and median \Av{} for all Perseus sources are 24.6 and 12,
respectively \citep{2006ApJ...638..293E}. The mean particle density of
the L1451 Bolocam sources ranges from 0.9~$\times$~10$^{5}$ to
1.5~$\times$~10$^{5}$~cm$^{-3}$, which is lower than the mean density
for all Perseus sources of 3.2~$\times$~10$^{5}$
\citep{2008ApJ...684.1240E}.  The kinetic temperature of gas within
the L1451 cores ranges from 9.1 to 10.3~K, which is lower than the
mean for all Perseus cores of 11.0 \citep{2008ApJS..175..509R}.  These
statistics complement the YSO statistics that suggest L1451 it is at
an earlier evolutionary epoch than Barnard~1 and NGC~1333.

The main science goals of our large-area, high-resolution, spectral
line observations of this young region are: 1) to quantify the dense
gas content of a cloud region possibly at the onset of star formation,
2) to determine whether complex, hierarchical structure formation
exists before the onset of star formation, as predicted by theories of
turbulence-driven star formation, 3) to better understand how natal
cloud material fragments on the pathway to star formation, by
quantifying the hierarchical similarities and differences between
sub-regions of L1451 with and without compact cores, 4) to estimate
the boundedness of the dense structures in young star forming regions,
and 5) to potentially discover new young cores.

The paper is organized as follows. Section~\ref{sec:ch3sec2} provides
an overview of CLASSy observations of L1451. Properties of the
L1451-mm continuum detection are in
Section~\ref{sec:ch3sec3}. Section~\ref{sec:ch3sec4} presents the
dense gas morphology using integrated intensity and channel maps, and
Section~\ref{sec:ch3sec5} presents the dense gas kinematic results
from spectral line fitting. A dendrogram analysis of the \HCO{}, HCN,
and \NtwoH{} data cubes is in Section~\ref{sec:ch3sec6}.
Section~\ref{sec:ch3sec7} shows how we calculate column density, dust
temperature, and extinction maps using \textit{Herschel} data, along
with a dendrogram analysis of the extinction
map. Section~\ref{sec:ch3sec8} discusses the current state of star
formation in L1451 using the spectral line data in combination with
the continuum data to further quantify physical and spatial properties
of structures in L1451. We summarize our key findings in
Section~\ref{sec:ch3sec9}.

\section{Observations}
\label{sec:ch3sec2}

\begin{figure}[!h]
\centering 
\includegraphics[scale=0.75]{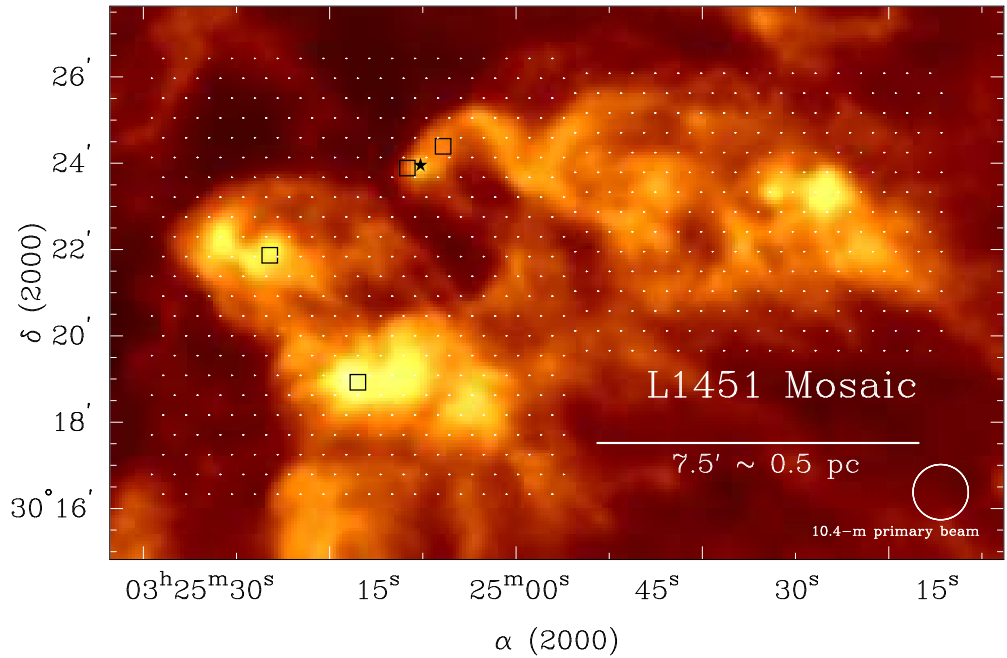}
\caption{\small A \textit{Herschel} image of the 250~$\mu$m emission
  (yellow is brighter emission, red is fainter emission) from L1451
  with the CLASSy mosaic pointing centers overlaid as white
  points. The spacing of the pointing centers is 31\arcsec{}, and our
  total area coverage is $\sim$150 square arcminutes. The locations of
  1.1~mm Bolocam sources \citep{2006ApJ...638..293E} and the L1451-mm
  compact continuum core \citep{2011ApJ...743..201P} are marked with
  black squares and a black star, respectively.}
\label{fig:mosaicpt}
\end{figure}

The details of CLASSy observations, calibration, and mapping are found
in Paper~I; specifics related to L1451 are summarized here. We
mosaicked a total area of $\sim$150 square arcminutes in CARMA23 mode,
which uses all 23 CARMA antennas.  The mosaic was made up of two
adjacent rectangles, containing a total of 673 individual pointings
with 31\arcsec{} spacing in a hexagonal grid (see
Figure~\ref{fig:mosaicpt}). The reference position of the mosaic is at
the center of the eastern rectangle:
$\alpha$=03$^{\textrm{\footnotesize h}}$25$^{\textrm{\footnotesize
    m}}$17$^{\textrm{\footnotesize s}}$,
$\delta$=30\degree21\arcmin23\arcsec (J2000). The L1451-mm core
\citep{2011ApJ...743..201P} is within the eastern rectangle. The
region was observed for 150 total hours, split between the DZ and EZ
configurations, which provide projected baselines from about
1--40~k$\lambda$ and 1--30~k$\lambda$, respectively, and a hybrid
array (DEZ) with baselines ranging from about 1--35~k$\lambda$. The
DEZ array was not used for the CLASSy observations presented in
Paper~I, and is the reason the synthesized beam for L1451 is slightly
larger than that for Barnard~1.  See Table~\ref{tbl:obssum} for a
summary of observing dates and calibrators. The mapped region covers
roughly 1.1~pc by 0.6~pc with about 1800~AU spatial resolution.

\begin{deluxetable}{l c c c c c }
\tabletypesize{\footnotesize} \tablecaption{Observation Summary}
\tablewidth{0pt} \setlength{\tabcolsep}{0.03in} \tablehead {
  \colhead{Array} & \colhead{Dates} & \colhead{Total Hours} &
  \colhead{Flux Cal.} & \colhead{Gain Cal.} & \colhead{Mean Flux (Jy)}
}
\startdata
DZ & October 2012 & 25 & Uranus & 3C84/3C111 & 18.6/2.6 \\
& April -- June 2013 & 19 & Uranus & 3C84/3C111 & 17.1/2.7 \\
DEZ & February 2013 & 31 & Uranus & 3C84/3C111 & 20.7/3.9 \\
EZ & August -- September 2012 & 25 & Uranus & 3C84/3C111 & 18.0/3.1 \\
& July -- August 2013 & 50 & Uranus & 3C84/3C111 & 17.5/4.2 \\
\enddata
\vspace{-0.5cm}
\label{tbl:obssum}
\end{deluxetable}

The correlator setup is summarized in Table~\ref{tbl:corrsum}.
\NtwoH{}, \HCO{}, and HCN ($J=1\rightarrow0$) were simultaneously
observed in 8~MHz bands, providing a velocity resolution of
0.16~\kms. We also used a 500~MHz band for continuum observations and
calibration. Data were inspected and calibrated using MIRIAD
\citep[Multichannel Image Reconstruction, Image Analysis and
  Display;][]{1995ASPC...77..433S} as described in Paper~I.  3C84 was
observed every 16 minutes for gain calibration; 3C111 was used for
gain calibration when 3C84 transited above 80 degrees
elevation. Uranus was observed for absolute flux calibration. The flux
of 3C84 varied between 16 and 21~Jy over the observing period, while
3C111 varied between 2.6 and 4.5~Jy. The uncertainty in absolute flux
calibration is about 10\%. We will only report statistical
uncertainties when quoting errors in measured values throughout the
paper.

\begin{deluxetable}{l c c c c c c c}
\tabletypesize{\tiny}
\tablecaption{Correlator Setup Summary}
\tablewidth{0pt}
\setlength{\tabcolsep}{0.03in}
\tablehead
{
 \colhead{Line} & \colhead{Rest Freq.} & \colhead{No. Chan.} & \colhead{Chan. Width} & \colhead{Vel. Coverage} & \colhead{Vel. Resolution} & \colhead{Chan. RMS} & \colhead{Synth. Beam$^{a}$} \\
 \colhead{} & \colhead{(GHz)} & \colhead{} & \colhead{(MHz)} & \colhead{(\kms)} & \colhead{(\kms)} & \colhead{(\Jybm)} & \colhead{} 

}
\startdata    
N$_{2}$H$^{+}$ & 93.173704$^{b}$ & 159 & 0.049 & 24.82 & 0.157 & 0.14 & 8.6\arcsec{}~$\times$~6.8\arcsec{} \\
Continuum      & 92.7947   &  47 & 10.4 & 1547 & 33.6 & 0.0013 & 9.2\arcsec{}~$\times$~6.6\arcsec \\
HCO$^{+}$      & 89.188518 & 159 & 0.049 & 25.92 & 0.164 & 0.12 & 8.8\arcsec{}~$\times$~7.1\arcsec{}\\
HCN            & 88.631847$^{c}$ & 159 & 0.049 & 26.10 & 0.165 & 0.12 & 8.9\arcsec{}~$\times$~7.2\arcsec{}\\
\enddata
\vspace{-0.5cm} \tablecomments{ $^{a}$The synthesized beam is slightly
  different for each pointing center, and MIRIAD calculates a
  synthesized beam for the full mosaic based on all of the pointings.
  $^{b}$The rest frequency of the band was set to the weighted mean
  frequency of the center three hyperfine components.  $^{c}$The rest
  frequency of the band was set to the frequency of the center
  hyperfine component. See $\textrm{http://splatalogue.net}$ for
  frequencies of the HCN and \NtwoH{} hyperfine components.}
\label{tbl:corrsum}
\end{deluxetable}

To create spectral-line images which fully recover emission at all
spatial scales, CARMA observed in single-dish mode during tracks with
stable atmospheric opacity. The OFF position for L1451 was
3.5\arcmin{} west and 13.7\arcmin{} south of the mosaic reference
position, at the location of a gap in $^{12}$CO and $^{13}$CO emission
to ensure no significant dense gas contribution. The single-dish data
from the 10.4-m dishes was calibrated in MIRIAD as described in
Paper~I.  The antenna temperature rms in the final single-dish cubes
was $\sim$0.02~K for all three molecules.  The spectral-line
interferometric and single-dish data were combined with {\tt mosmem},
a maximum entropy joint deconvolution algorithm in MIRIAD. The noise
levels and synthesized beams for the final data cubes are given in
Table~\ref{tbl:corrsum}. The rms noise in these lines correspond to
brightness temperature rms of 0.34~K for \NtwoH{} and 0.30~K for
\HCO{} and HCN.

We created a 3~mm continuum map with the interferometric data from the
500~MHz window. The rms in the continuum map is $\sim$1.3 m\Jybm{}
with a synthesized beam of
9.2\arcsec{}~$\times$~6.6\arcsec. Single-dish continuum data can not
be acquired at CARMA.

\section{Continuum Results}
\label{sec:ch3sec3}

\begin{figure}[!h]
\centering \includegraphics[scale=0.55]{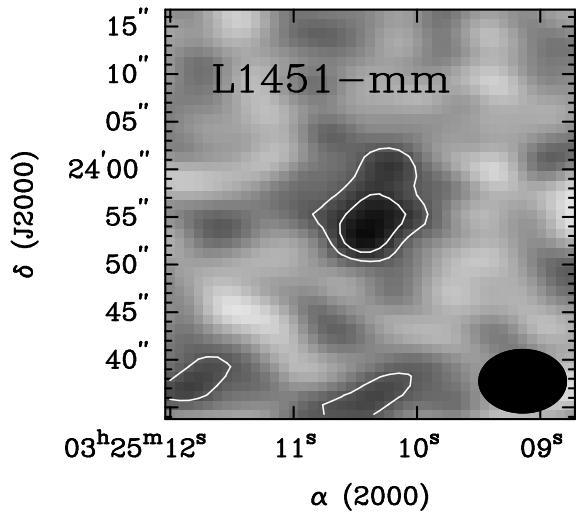}
\caption{\small The single continuum detection in our field. The
  synthesized beam is 9.2\arcsec{}~$\times$~6.6\arcsec, and the
  1$\sigma$ sensitivity is 1.3~m\Jybm. The contour levels are 2, 3
  times 1$\sigma$; no negative contours are present.}
\label{fig:cont}
\end{figure}

We detected no compact continuum sources above the 5$\sigma$ level of
the 3~mm continuum map. One source was detected above 3$\sigma$ that
could be confirmed with other observations; L1451-mm
\citep{2011ApJ...743..201P} is detected at 4$\sigma$ with
5.2~m\Jybm. Figure~\ref{fig:cont} shows the 3~mm continuum image
toward L1451-mm. The position, peak brightness, and lower-limit mass
for our detection were calculated following the prescription described
in Paper~I and are listed in Table~\ref{tbl:cont}. The position and
peak brightness agree with CARMA 3~mm measurements from
\citet{2011ApJ...743..201P}, which had a $\sim$5\arcsec{} synthesized
beam. Our image shows a possible secondary peak to the north of the
brightest emission. However, this secondary peak is only within the
2--3$\sigma$ contours and does not appear in the higher sensitivity
observations of \citet{2011ApJ...743..201P}.
\citet{2011ApJ...743..201P} detected a low-velocity CO
($J=2\rightarrow1$) outflow in this area; we do not detect any HCN or
\HCO{} outflow emission near this source.

\begin{deluxetable}{l c c c }
\tabletypesize{\footnotesize} \tablecaption{Observed Properties of Continuum Detection} \tablewidth{0pt} \setlength{\tabcolsep}{0.03in}
\tablehead{ 
\colhead{Source} & \colhead{{Position}} & \colhead{{Pk. Bright.}} & \colhead{{Mass}} \\
\colhead{Name} & \colhead{{(h:m:s, d:\arcmin:\arcsec)}} & \colhead{{(m\Jybm)}} & \colhead{{(\Msun)}}\\
\colhead{(1)} & \colhead{{(2)}} & \colhead{{(3)}} & \colhead{{(4)}} 
}
\startdata 
L1451-mm & 03:25:10.38, +30:23:55.9  & 5.2 $\pm$ 1.3 & 0.10~$\pm$~0.03 \\
\enddata
\vspace{-0.5cm} \tiny \tablecomments{ (3) Peak brightness, (4)
  Lower-limit mass using the peak brightness and assumptions outlined
  in Paper~I.}
\label{tbl:cont}
\end{deluxetable}

\section{Morphology of Dense Molecular Gas}
\label{sec:ch3sec4}

\subsection{Integrated Intensity Emission}
\label{sec:ch3sec4sub1}

\begin{figure}[!h]
\centering \includegraphics[scale=0.23]{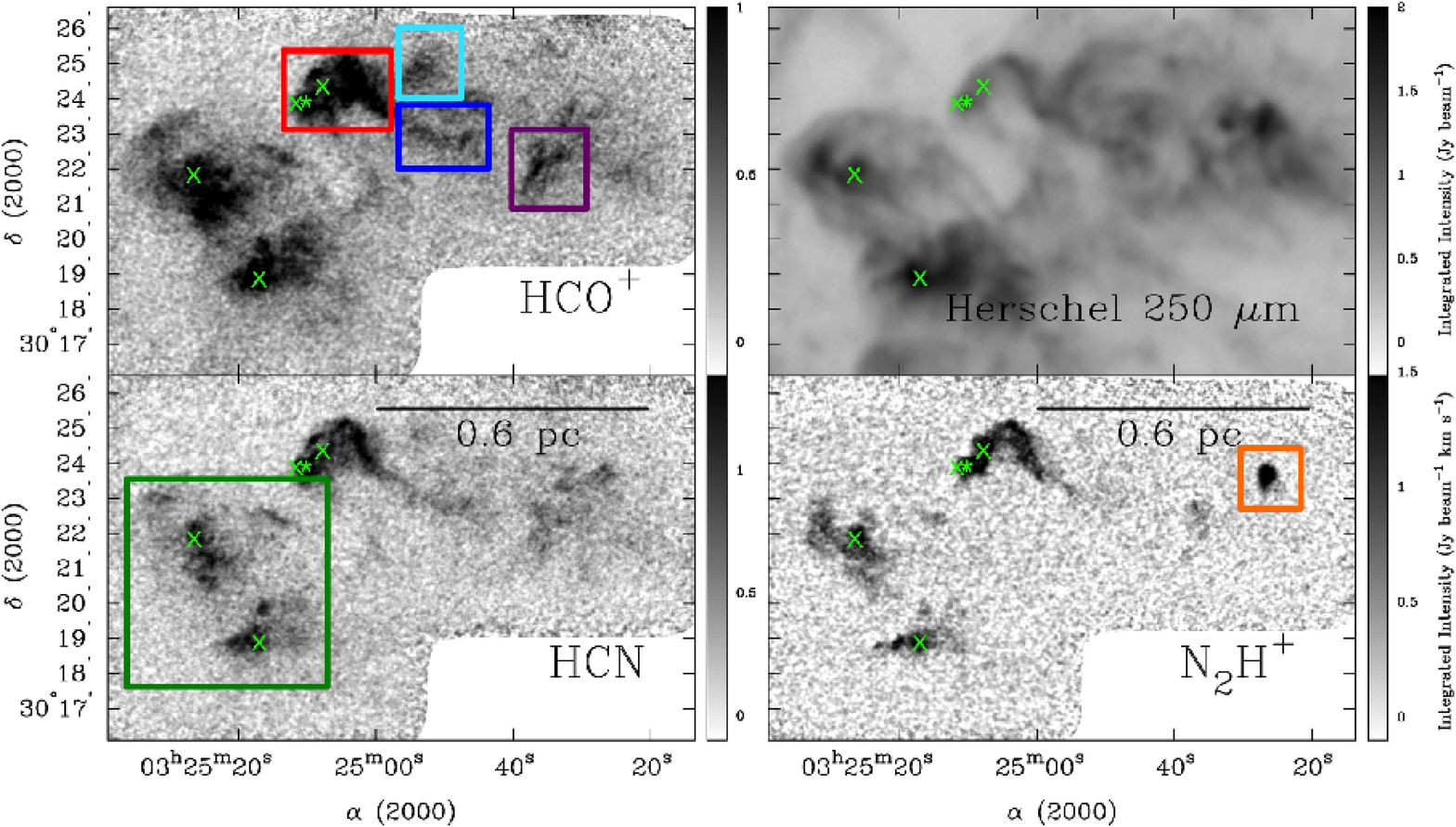}
\caption{ \small Integrated intensity maps of \HCO{}, HCN, and \NtwoH
  ($J=1\rightarrow0$) emission toward L1451, along with a
  \textit{Herschel} 250~$\mu$m map of the region. \HCO{} emission was
  integrated from 5.316 to 3.018~\kms. HCN emission was integrated
  over all three hyperfine components from 9.945 to 7.962~\kms, 5.154
  to 2.841~\kms, and $-$2.115 to $-$3.932~\kms.  \NtwoH{} emission was
  integrated over all seven hyperfine components over velocity ranges
  from 11.542 to 8.714~\kms, 5.729 to 2.744~\kms, and $-$3.383 to
  $-$4.639~\kms. The rms of the maps are 0.08, 0.13, and
  0.16~\Jybmkms{} for \HCO{}, HCN, and \NtwoH, respectively.  The four
  Bolocam 1.1~mm sources in this region are marked with ``x'' symbols,
  and the L1451-mm compact continuum core is marked with an
  asterisk. Colored rectangles show the locations of qualitative
  features discussed in Section~\ref{sec:ch3sec4sub1}.}
\label{fig:birdseye}
\end{figure}

Figure~\ref{fig:birdseye} shows integrated intensity maps for \HCO{},
HCN, and \NtwoH{} ($J=1\rightarrow0$) ($\sim$8\arcsec{} angular
resolution), along with a \textit{Herschel} 250~$\mu$m image
(18.1\arcsec{} angular resolution) for a visual comparison between the
dense gas and dust emission.  The line maps were integrated over all
channels with identifiable emission. The locations of the four Bolocam
1.1~mm sources \citep{2006ApJ...638..293E} and the one compact
continuum core in L1451 are marked on each image of
Figure~\ref{fig:birdseye}. Four of the five sources are located near
peaks of dense gas and dust emission. While the molecules and dust are
tracing similar features around those sources, the exact morphological
details vary. Below, we describe the qualitative emission features,
and refer to the colored rectangles in Figure~\ref{fig:birdseye} for
reference.

All tracers show a curved structure surrounding L1451-mm and the two
nearby Bolocam sources (see red rectangle in
Figure~\ref{fig:birdseye}), with a peak of integrated emission at the
location of L1451-mm. The southwestern edge of the curved structure
has a stream of emission that extends further to the southwest (see
dark blue rectangle in upper left panel of Figure~\ref{fig:birdseye}),
which can be seen in all the maps, though it extends furthest in the
\HCO{}, HCN, and dust maps. The two other Bolocam sources to the far
east of L1451-mm are surrounded by significant molecular gas structure
(see green rectangle in lower left panel). The \HCO{} emission has the
largest spatial extent in this region.

The integrated emission in the three lines is less similar across the
western half of L1451 compared to the eastern half. There is a strong,
condensed \NtwoH{} source (see orange rectangle in lower right panel)
that does not appear strongly in the HCN or \HCO{} maps, but that does
correspond to a peak of emission in the dust map (see
Section~\ref{sec:ch3sec8sub3} for more details on this source).  The
strongest \HCO{} feature in the western half of the map has a weaker
counterpart in the HCN map (see purple rectangle), which appears even
weaker in \NtwoH{}. Finally, there is \HCO{} emission to the northwest
of the curved structure (see cyan rectangle) that closely mimics dust
emission in that region; this emission is weakly detected in HCN, but
not in \NtwoH{}. Since the $J=1\rightarrow0$ transition of \HCO{}
traces densities about an order of magnitude lower than the other two
molecules \citep{2015PASP..127..299S}, the regions with strong \HCO{}
and weak HCN and \NtwoH{} are likely at lower density compared to
regions where all the molecules have strong emission.

\subsection{Channel Emission}
\label{sec:ch3sec4sub2}
\begin{figure}[!h]
\centering \includegraphics[scale=0.9]{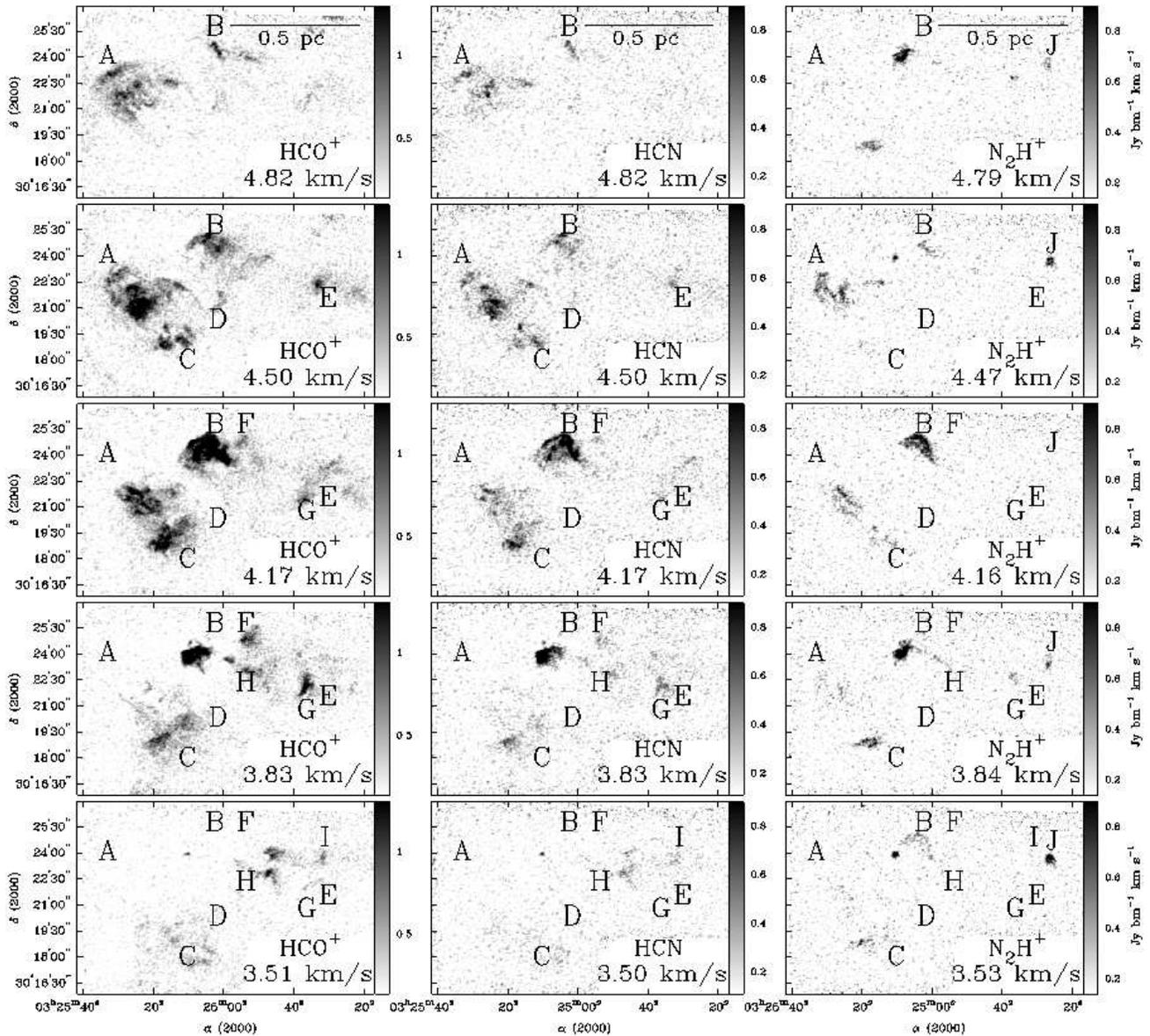}
\caption{\small \textit{Left}: Five \HCO{} channels maps, with
  two-channel spacing. The rms in each channel is 0.12~\Jybm{}, and
  the color intensity ranges from 0.12--1.3~\Jybm. Features discussed
  in the text are identified with a letter in the first channel they
  appear. \textit{Center}: Five HCN channels maps. The rms in each
  channel is 0.12~\Jybm{}, and the color intensity ranges from
  0.12--0.9~\Jybm. Feature labels from the \HCO{} maps are overplotted
  in the same locations for aid in comparing the emission across
  molecules.  \textit{Right}: Five \NtwoH{} channel maps. Note that
  several features appear twice across the channels due to the
  hyperfine structure of \NtwoH{}.  The rms in each channel map is
  0.14~\Jybm{}, and the color intensity ranges from
  0.14--0.9~\Jybm. Feature labels from the \HCO{} maps are overplotted
  in the same locations for aid in comparing the emission across
  molecules. Feature~J (also referred to as L1451-west) only appears
  in \NtwoH{}, so it is only labeled in this figure.}
\label{fig:molchans}
\end{figure}

Figure~\ref{fig:molchans} shows channel maps of \HCO{}, HCN, and
\NtwoH{} highlighting the bulk of the emission, which occurs from
$\sim$4.8 to 3.5~\kms{}, with 2-channel spacing (e.g., we skip the
4.66~\kms{} channel between the 4.82~\kms{} and 4.50~\kms{} channels
of \HCO{}).  In general, it is clear from the channel maps in
Figure~\ref{fig:molchans} that strong \HCO{} emission is more
widespread compared to HCN emission, and particularly \NtwoH{}
emission (as was also evident in Figure~\ref{fig:birdseye}).  We label
qualitative features in the \HCO{} channel maps, from A through I, in
the order that they appear in velocity space (with eastern sources
being labeled before western sources). We then place those same labels
on the HCN and \NtwoH{} maps to aid a qualitative comparison of dense
gas features, given below.

Features A and C in the eastern half of the map are traced with all
molecules, with varying strength. In Figure~\ref{fig:molchans},
Feature A appears strongly in \HCO{} at 4.82~\kms{}, faintly at
4.82~\kms{} in HCN, and not until 4.47~\kms{} in \NtwoH{}. Even when
it finally appears as \NtwoH{} emission, the emission is much more
spatially concentrated than what we observe in \HCO{} and HCN. This
spatial concentration is likely because the \NtwoH{}
($J=1\rightarrow0$) line traces higher-density, colder material
compared to \HCO{} and HCN ($J=1\rightarrow0$) lines
\citep{2015PASP..127..299S}. A 1.1~mm Bolocam core is located within
Feature A, and corresponds with a peak in the \NtwoH{}
emission. Feature A moves northeast to southwest from 4.82~\kms{} to
lower-velocity channels as Feature C appears to its
southwest. Features A and C are possibly part of the same larger-scale
structure, which will be explored in the next section when we analyze
the kinematics of this region. Like Feature A, Feature C is more
spatially concentrated in \NtwoH{}, and contains a 1.1~mm Bolocam
source at an \NtwoH{} peak of emission.

Feature B contains the L1451-mm compact continuum source. It appears
strongly in all molecules, though it contains a prominent ridge of
emission at 4.82~\kms{} in \HCO{} and HCN that does not appear in
\NtwoH{} near that velocity.

Features D, E, F, G, H, and I are all identified based on the \HCO{}
emission. HCN emission appears weakly toward all features seen with
\HCO{}, while \NtwoH{} only shows faint emission in one channel for
Features G and H at 3.84~\kms{}. The descriptions below are based on
\HCO{}.  Feature D appears to the west of Features A and C, and to the
south of Feature B, seen at 4.50~\kms{} as an elliptical feature.
Feature E appears as a prominent, round emission feature at
4.50~\kms{}, with more extended emission in channels surrounding the
4.50~\kms{} peak of emission.  Feature F is emission that starts just
to the northwest of Feature B at 4.17~\kms{}, peaks at 3.83~\kms{},
and then appears to get fainter while extending to the southwest at
the unshown 3.67~\kms{} channel, while then brightening to the
southwest at 3.51~\kms{}.  Feature G emission peaks at 3.83~\kms{},
and appears as a stream of emission to the east of Feature~E.  Feature
H is a streamer to the southwest of Feature B and the south of Feature
F, which first appears at 3.83~\kms{}. It persists at 3.51~\kms{} and
more faintly extends into two lower-velocity channels not shown in the
figure.  Feature~I first appears at 3.51~\kms{}, brightens in two
lower-velocity channels not shown in the figure, and is not detectable
at velocities lower than that.

Feature J, referred to as L1451-west in the rest of the paper, is
relatively round emission that only appears strongly in the \NtwoH{}
data. It first appears at 4.79~\kms{} in Figure~\ref{fig:molchans},
and is visible across a total of four velocity channels per hyperfine
component---the structure can be seen repeating for another hyperfine
component, starting in the 3.84~\kms{} channel. We discuss details of
the structure and kinematics of L1451-west in
Section~\ref{sec:ch3sec8sub3}.

\begin{figure}[!h]
\centering \includegraphics[scale=0.5]{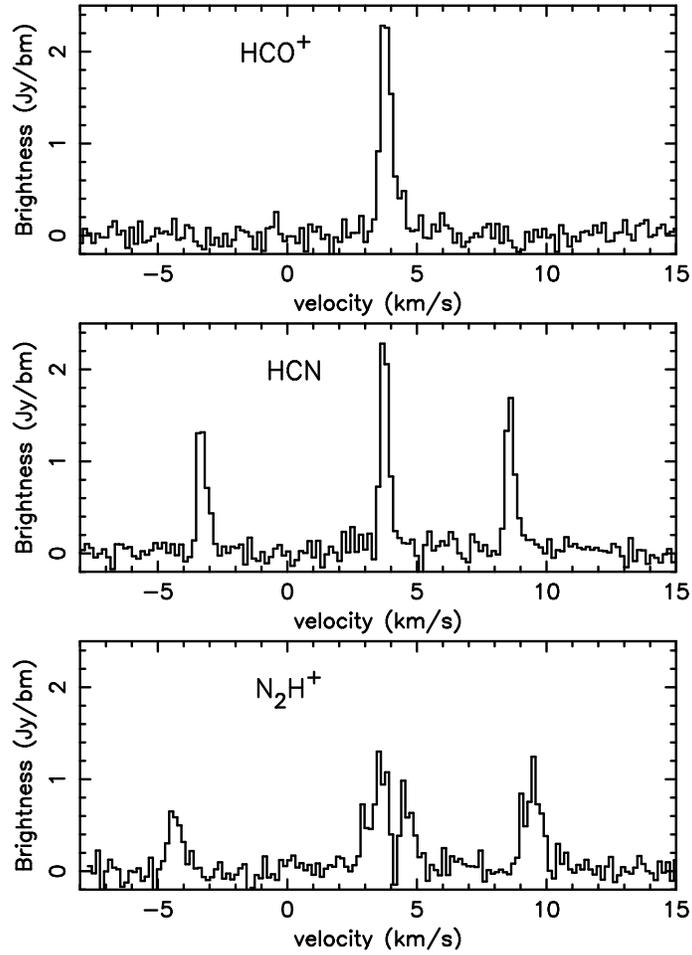}
\caption{\small Example \HCO{}, HCN, and \NtwoH{} spectra from the
  location of L1451-mm reported in Table~\ref{tbl:cont}. Spectra are
  averaged over one synthesized beam, and the conversion factors from
  \Jybm{} to Kelvin are reported in the text.}
\label{fig:newspec}
\end{figure}

We detect no HCN or \HCO{} outflow emission in any channel, which
suggests that L1451 is a young region with little to no protostellar
activity. Figure~\ref{fig:newspec} show an example spectrum for each
molecule from the location of L1451-mm within a single synthesized
beam. The conversions from \Jybm{} to K for these data are
2.47~K/\Jybm{}, 2.46~K/\Jybm{}, and 2.42~K/\Jybm{} for \HCO{}, HCN,
and \NtwoH{}, respectively.

A small fraction of \HCO{} and HCN spectra across L1451 show double
peaks with a variety of line-shape characteristics---we estimate that
$\sim$3\% of cloud locations show double-peak features. We cannot
determine the absolute cause of the double peaks without H$^{13}$CN
and H$^{13}$CO$^{+}$ ($J=1\rightarrow0$) observations at each cloud
location, so analyzing these features is beyond the scope of this
paper. However, the most likely scenario is self-absorption from a
foreground screen of lower-density gas with a significant \HCO{} and
HCN ($J=1\rightarrow0$) population. An infall scenario can be ruled
out in many locations since infall predicts a stronger blue peak
\citep{1999ARA&A..37..311E} while we often observe stronger red peaks;
a scenario with two-components along the same line-of-sight can likely
be ruled out in many locations where the \HCO{} and HCN spectra do not
both show double-peaks.

\clearpage

\section{Kinematics of Dense Molecular Gas}
\label{sec:ch3sec5}

We fitted the molecular line emission presented in
Section~\ref{sec:ch3sec4} with Gaussians using the method described in
Paper~I, and present the centroid velocity and line-of-sight velocity
dispersion maps here. The seven resolvable hyperfine components of the
\NtwoH{} ($J=1\rightarrow0$) line, and the three hyperfine components
of the HCN ($J=1\rightarrow0$) line, are simultaneously fit assuming
the same velocity dispersion and excitation conditions for each
component. \HCO{} ($J=1\rightarrow0$) has no hyperfine splitting and
is fit as a single Gaussian component.

We fit all the HCN and \HCO{} spectra with a single component across
the entire field, even though about 3\% of the spectra show evidence
of double-peaks. To estimate the impact of fitting a double-peaked
spectrum with a single component, we extrapolated several
double-peaked spectra to a single peak, fit those single-peak spectra,
and measured the full width at half of the extrapolated single-peak
spectra. From those fits, we estimate that the velocity dispersions
derived from the single-component fits to double-peaked spectra are
overestimated by $\sim$10\%. Considering that only about 3\% of the
spectra have this 10\% overestimation, the double-peaks have a
negligible impact on the results presented in this paper.

Although a low-velocity outflow from L1451-mm was previously detected
in CO ($J=2\rightarrow1$) by \citet{2011ApJ...743..201P}, we do not
detect any outflow signatures in our HCN or \HCO{} ($J=1\rightarrow0$)
observations as we did in other CLASSy fields. Therefore, there are no
line broadening effects from outflows that are impacting the line
profile fits.

\begin{figure}[!h]
\centering \includegraphics[scale=0.85]{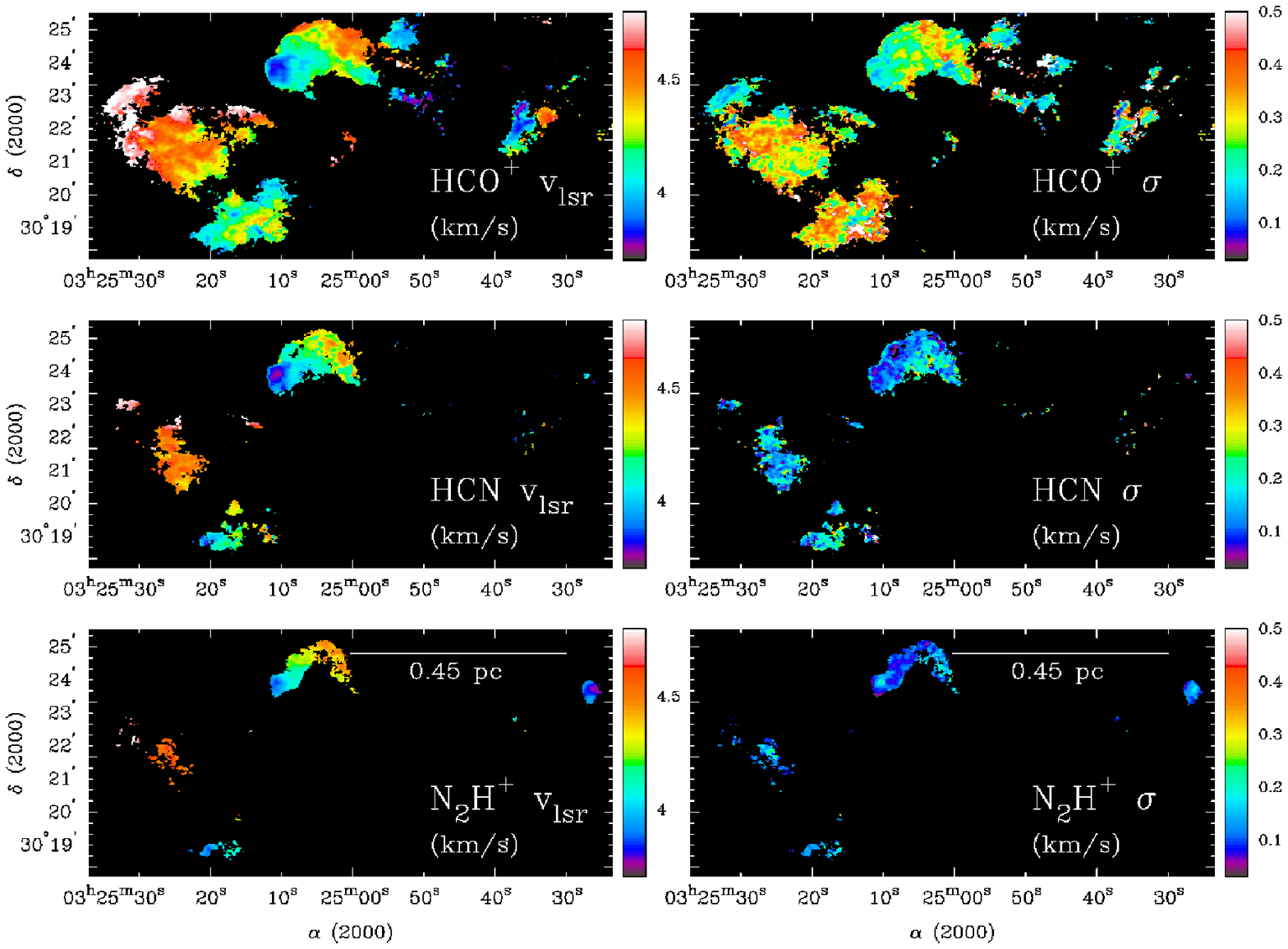}
\caption{\small Kinematics of dense gas in L1451. \textit{Left:}
  Centroid velocity (\kms{}) maps of \HCO, HCN, and \NtwoH{}
  ($J=1\rightarrow0$) emission, from top to bottom. \textit{Right:}
  Line-of-sight velocity dispersion (FWHM/2.355 in \kms{}) maps of
  \HCO{}, HCN, and \NtwoH{} ($J=1\rightarrow0$) emission, from top to
  bottom. We masked these maps to visualize only statistically robust
  kinematic results (see Section~\ref{sec:ch3sec5} text). The color
  scales are the same across molecules. }
\label{fig:kin}
\end{figure}

In Figure~\ref{fig:kin}, we plot the fitted centroid velocity and
velocity dispersion where: 1) the integrated intensity is greater than
or equal to four times the rms of the integrated intensity map, and 2)
the peak signal-to-noise of the spectrum is greater than five. We will
use these kinematic maps in the following sections in order to
interpret the hierarchical and turbulent nature of the region. Here we
list some general features of interest pertaining to the maps:

\begin{enumerate}

\item \HCO{} has systematically larger line-of-sight velocity
  dispersion compared to HCN and \NtwoH{}. The mean and standard
  deviation of the velocity dispersions across the maps are
  0.29~$\pm$~0.10, 0.16~$\pm$~0.07, and 0.12~$\pm$~0.04~\kms, for
  \HCO{}, HCN, and \NtwoH{}, respectively. The observed mean velocity
  dispersions above should be compared to the isothermal sound speed
  of the mean gas particles in the cloud for determining the
  turbulence in the observed gas. Note that the thermal velocity
  dispersion of the mean gas particle in the cloud is different from
  the thermal velocity dispersion of an individual molecule.  If we
  assume that the typical temperature in this region is 10~K based on
  ammonia observations of Bolocam cores \citep{2008ApJS..175..509R},
  then the isothermal sound speed between the mean gas particles would
  be $\sim$0.2~\kms{} (assuming molecular weight per free particle of
  2.33), while the isothermal sound speed between individual \NtwoH{}
  particles would be $\sim$0.05~\kms.

  The \NtwoH{} velocity dispersions are subsonic everywhere, the HCN
  velocity dispersions are subsonic in most cloud locations, and the
  \HCO{} velocity dispersions are transsonic to supersonic in most
  cloud locations. Note that even though \NtwoH{} and HCN are subsonic
  in many areas of L1451, they are not exhibiting purely thermal
  velocity dispersions, which would be $\sim$0.05~\kms{} for 10~K gas.
  The $J=1\rightarrow0$ line of \HCO{} traces densities about an order
  of magnitude lower than that of HCN and \NtwoH{}
  \citep[see][]{2015PASP..127..299S}. Therefore, we are likely
  observing the trend from supersonic to subsonic gas motions as gas
  goes from the larger, less-dense scales traced by \HCO{} to the
  smaller, more-dense scales traced by HCN and \NtwoH{}. This is
  expected in a turbulent medium, where velocity dispersion scales
  proportionally with size \citep{2007ARAA..45..565M}.

\item All three molecules are tracing gas with centroid velocities
  ranging from $\sim$3.8--4.7~\kms. However, the \HCO{} gas extends to
  lower velocities, due to the gas in the Feature~H streamer, which
  appears in the HCN maps, but is not strong enough to provide
  reliable kinematic measurements. \HCO{} also extends to higher
  velocities, due to strong gas emission from the northeast part of
  Feature A, which is noticeable in the first channel of
  Figure~\ref{fig:molchans}.

\item The \HCO{} centroid velocities for Features A and C show a
  gradient from northeast to southwest. It is possible that this is a
  large, rotating piece of dense gas, which is fragmenting into denser
  components (e.g., the Bolocam 1~mm sources). It is also possible
  that the redshifted northeast section and the blueshifted southwest
  section represent independent components in the turbulent medium, or
  that they are merely projected along the same line of
  sight. Observations of optically thin tracers, and tracers of
  lower-density, larger-scale material are needed to help distinguish
  between these scenarios.

\item The gas in the eastern half of Feature~B shows a velocity
  gradient along the length of the feature. It is most blueshifted at
  the southeastern end near the L1451-mm core, and becomes
  increasingly redshifted further away to the northwest.  The gas
  immediately surrounding the L1451-mm compact continuum core has a
  centroid velocity pattern that is consistent with rotation
  \citep{2011ApJ...743..201P}, and has velocity dispersions that
  increase toward the core center. L1451-west \NtwoH{} velocity
  dispersions also peak at the core center, and we compare these two
  sources in detail in Section~\ref{sec:ch3sec8sub3}.

\item Our measurements of \NtwoH{} centroid velocity and velocity
  dispersion towards L1451-mm agree well with the results in
  \citet{2011ApJ...743..201P}, in terms of absolute values measured
  and gradients across the source.

\end{enumerate}

\section{Dendrogram Analysis of Molecular Emission}
\label{sec:ch3sec6}   

\subsection{The Non-binary Dendrograms}
\label{sec:ch3sec6sub1}   
We qualitatively described the dense gas morphology of L1451 in
Section~\ref{sec:ch3sec4}. Here we quantitatively identify dense gas
structures from the three position-position-velocity (PPV) cubes and
study the hierarchical nature of L1451 with the non-binary dendrogram
algorithm described in Paper~I.

A dendrogram algorithm identifies emission peaks in a dataset and
keeps track of how those peaks merge together at lower emission
levels. This method of identifying and tracking emission structures is
advantageous compared to a watershed object-identification algorithm,
such as CLUMPFIND~\citep{1994ApJ...428..693W}, when the science goals
include understanding how the morphology and kinematics of
star-forming gas connects from large to small scales. A full
discussion of the most widely used dendrogram algorithm applied to
astronomical data can be found in \citet{2008ApJ...679.1338R}; details
of dendrograms and our non-binary version of the dendrogram algorithm
are found in the appendices of Paper~I, along with a comparison with
the results from using a more standard clump-finding algorithm on our
CLASSy data.

We ran our non-binary dendrogram algorithm on the \HCO{} emission, the
emission from the strongest HCN hyperfine line, and the emission from
the strongest \NtwoH{} hyperfine line. For other CLASSy regions, we
limited our analysis to \NtwoH{} emission because the \HCO{} and HCN
lines were complicated by protostellar outflows and severe
self-absorption effects not seen in L1451. Also, for other CLASSy
regions, we ran the algorithm on the \NtwoH{} hyperfine component most
isolated in velocity space because the strongest hyperfine component
was often not spectrally resolved from adjacent components.  The
\NtwoH{} hyperfine components in L1451 are resolved in all locations,
letting us use the strongest component for our dendrogram analysis. A
caveat is that some bluer emission from the higher-velocity, adjacent
hyperfine component, and some redder emission from the lower-velocity,
adjacent hyperfine component appear in the channels of the strongest
component. Since there is no blending of hyperfine emission at the
same location within the cloud, we masked the emission from the
adjacent hyperfine components in the individual channels of the
strongest hyperfine component in which they appeared.  As an example,
L1451-west (Feature~J) appears in much bluer channels compared to
Feature~A, so L1451-west emission from the higher-velocity, adjacent
hyperfine component also appears in the reddest channel of the
strongest hyperfine component that shows Feature~A.  For this example,
we masked out L1451-west emission from this red channel of the
strongest component.

We ran the algorithm with similar parameters used for the Barnard~1
analysis described in Paper~I, while following the prescription
presented in Appendix~\ref{app:appendixC} for comparing data cubes
with different noise levels. The critical algorithm inputs and
parameters were: (1) a masked input data cube with all pixels greater
than or equal to 4$\sigma$ intensity, along with pixels adjacent to
the initial selection that are at least 2.5$\sigma$ intensity, where
$\sigma$ is the rms level of the given data cube, (2) a set of local
maxima greater than or equal to all their neighbors in 10\arcsec{} by
10\arcsec{} by three channel (0.94~\kms) spatial-velocity pixels to
act as potential dendrogram leaves, (3) a requirement that a local
maximum peak at least 2-$\sigma_{n}$ above the intensity where it
merges with another local maximum for a structure to be considered a
leaf (referred to as the ``{\tt minheight}'' parameter below and in
Appendix~\ref{app:appendixC}), where $\sigma_{n}$ is the rms level of
the noisiest data cube (\NtwoH{} at $\sim$0.14~\Jybm{}, in this case),
and (4) a requirement of at least three synthesized beams of
spatial-velocity pixels for a structure to be considered a leaf
(referred to as the ``{\tt minpixel}'' parameter below and in
Appendix~\ref{app:appendixC}).  The {\tt minheight} and {\tt minpixel}
parameters act to prevent noise features from being identified as
dendrogram leaves. Branching steps are restricted to integer values of
the 1-$\sigma_{n}$ sensitivity of the data (referred to as the ``{\tt
  stepsize}'' parameter below and in Appendix~\ref{app:appendixC}) for
our non-binary dendrograms when comparing datasets with different
noise-levels. Appendix~\ref{app:appendixC} shows that using uniform
{\tt minheight} and {\tt stepsize} allows a comparison of dendrogram
properties that minimizes the impact of noise-level differences
between data cubes.

\begin{figure}[!h]
\centering \includegraphics[scale=0.13]{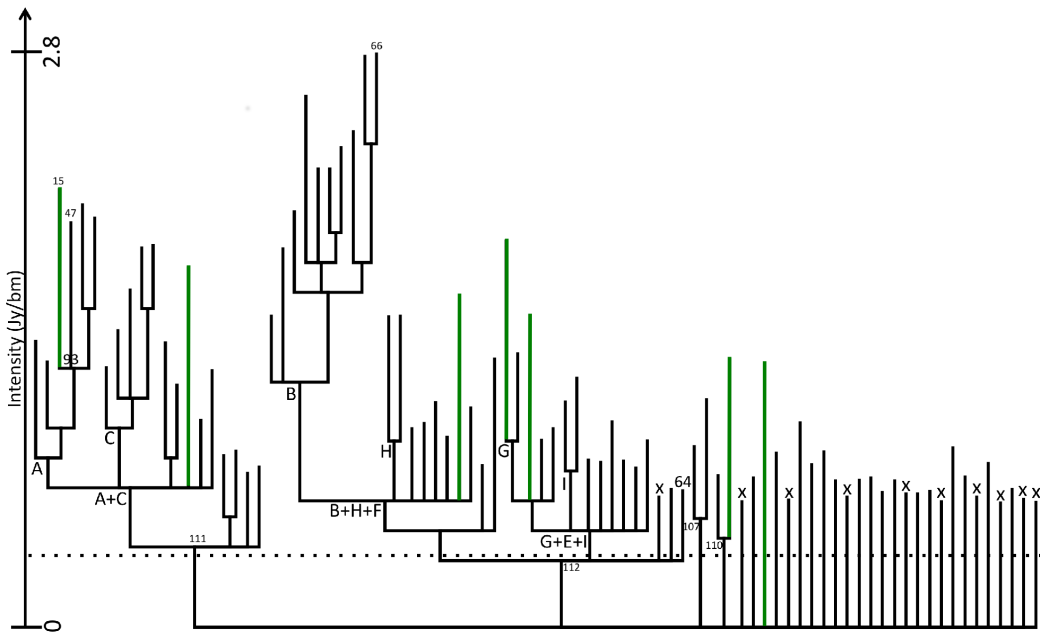}
\caption{\small The \HCO{} non-binary dendrogram for L1451. The
  vertical axis represents the \Jybm{} intensity for a given location
  within the gas hierarchy. The horizontal axis is ordered so that
  features identified in Section~\ref{sec:ch3sec4} are generally
  ordered from east to west. Leaves and branches discussed in the text
  are labeled with their numerical identifier, and branches associated
  with major features from Section~\ref{sec:ch3sec4} are marked with
  the feature letter.  The leaves colored green peak at least
  6-$\sigma_{n}$ above their first merge level. The horizontal dotted
  line represents the 2.5-$\sigma_{n}$ intensity cut above which we
  calculate tree statistics \textit{when comparing dendrograms made
    from data with different noise levels}. The leaves labeled ``x''
  are discarded from the calculation of tree statistics when comparing
  the dendrograms of different molecules observed with different noise
  levels, but they are used when studying the structure of an
  individual dendrogram.}
\label{fig:hcopdendrogram}
\end{figure}

\begin{figure}[!h]
\centering \includegraphics[scale=0.15]{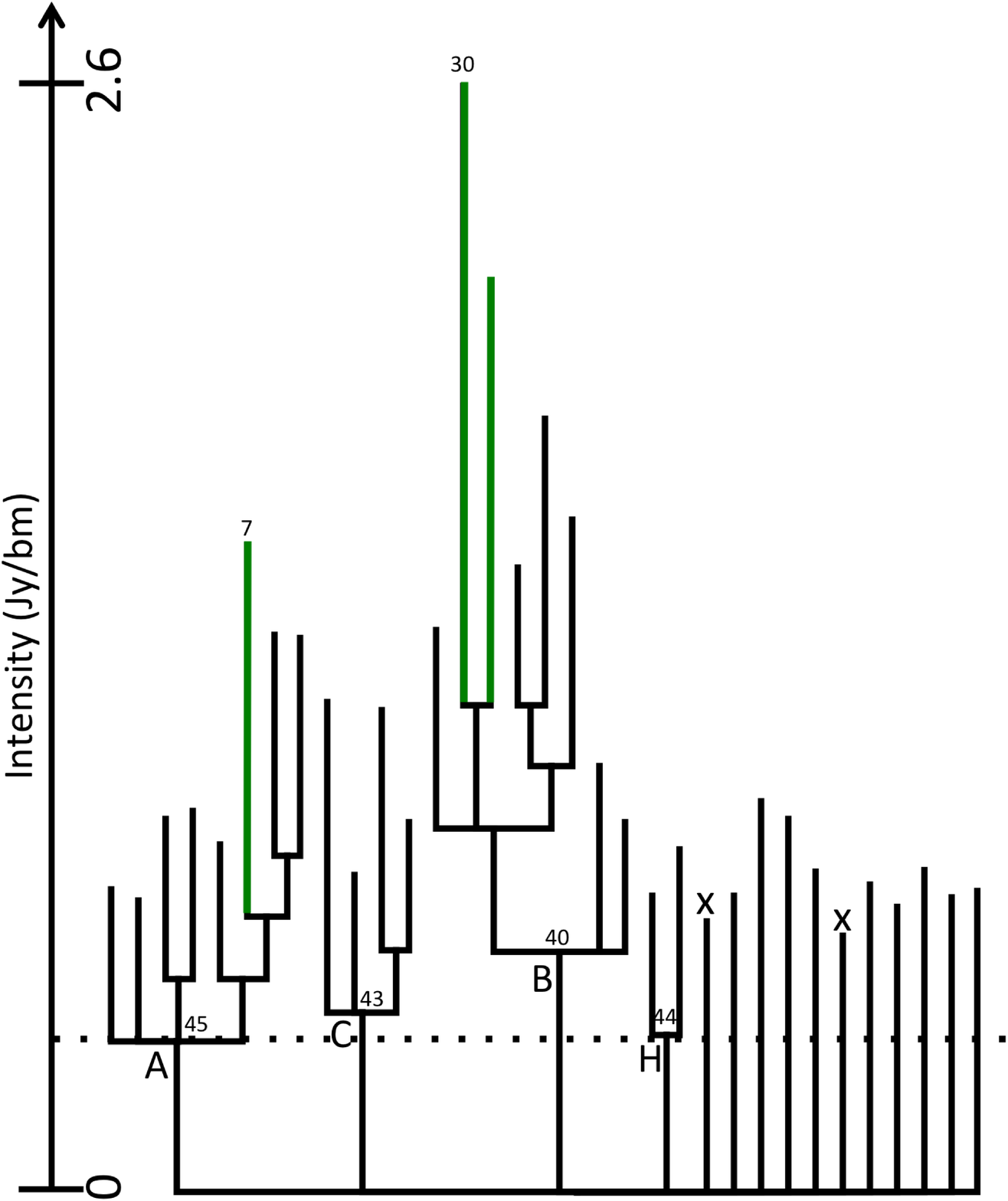}
\caption{\small Same as Figure~\ref{fig:hcopdendrogram}, but for the
  HCN non-binary dendrogram.}
\label{fig:hcndendrogram}
\end{figure}

\begin{figure}[!h]
\centering \includegraphics[scale=0.15]{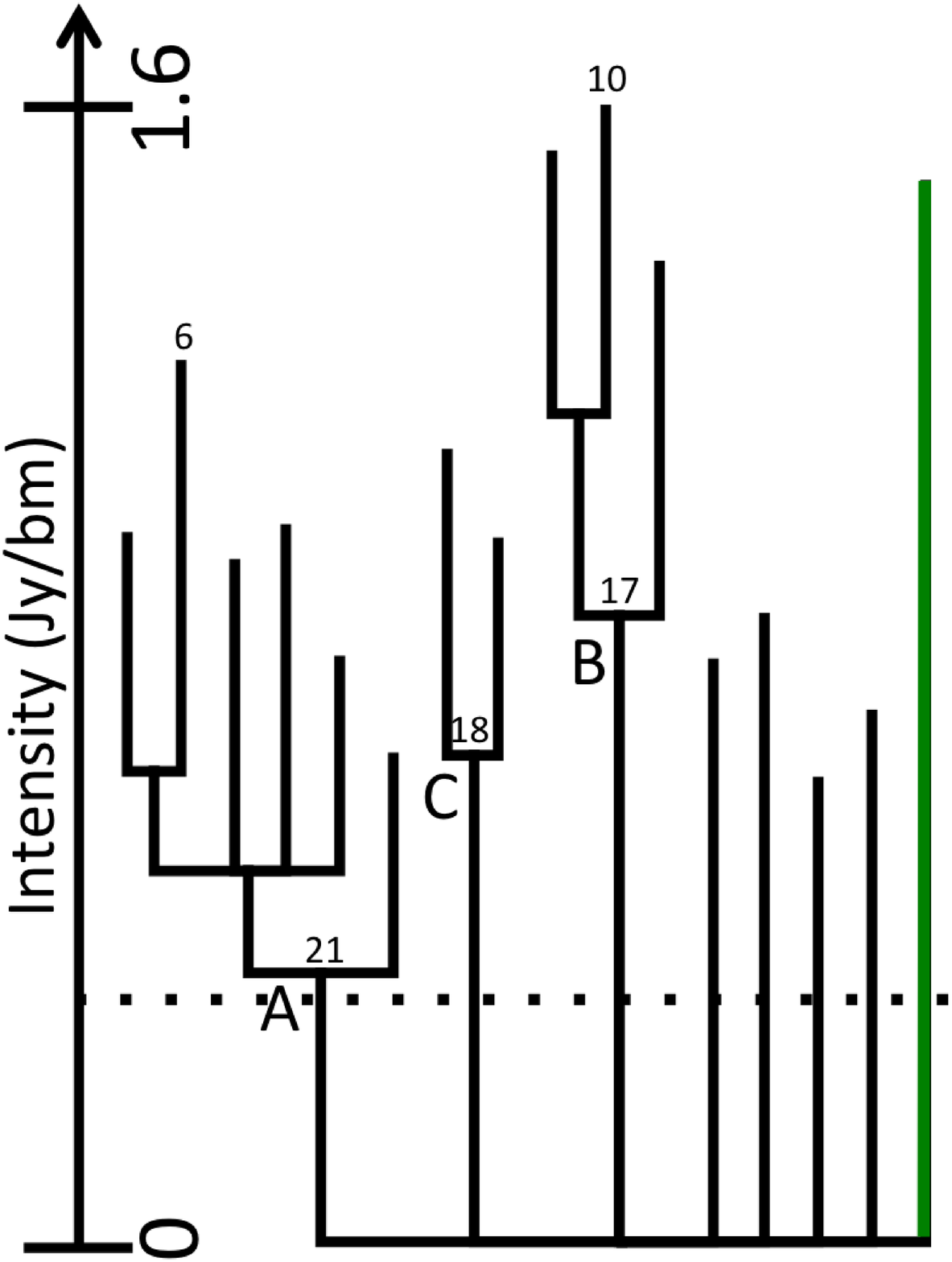}
\caption{\small Same as Figure~\ref{fig:hcopdendrogram}, but for the
  \NtwoH{} non-binary dendrogram.}
\label{fig:n2hpdendrogram}
\end{figure}

Figures~\ref{fig:hcopdendrogram}, \ref{fig:hcndendrogram}, and
\ref{fig:n2hpdendrogram} show \HCO{}, HCN, and \NtwoH{} non-binary
dendrograms for L1451, respectively. The vertical axis of the
dendrograms represent the intensity range of the pixels belonging to a
leaf or branch. The horizontal axis is arranged with the major
features identified in Figure~\ref{fig:molchans} progressing from
east-to-west; we label certain branches that are associated with major
features, and we provide the numeric label for structures referred to
in the upcoming discussion. The isolated leaves are presented in
numerical order. The horizontal dotted line in each dendrogram
represents an intensity cut at 2.5-$\sigma_{n}$ that aids in
cross-comparison of dendrogram statistics (see
Appendix~\ref{app:appendixC}) and is discussed in
Section~\ref{sec:ch3sec6sub3}.

The \HCO{} dendrogram contains the largest number of structures, with
86 leaves and 27 branches. The HCN dendrogram contains 33 leaves and
13 branches, while the \NtwoH{} dendrogram only contains 16 leaves and
6 branches. Leaves that peak at least 6-$\sigma_{n}$ in intensity
above the branch they merge directly into are colored green and
referred to as high-contrast leaves\footnote{We will use the
  definition of high-contrast leaves, first introduced in Paper~I, as
  a way of comparing the properties of strong leaves across CLASSy
  clouds in future papers.}. The strongest leaf for every molecule is
at or near the location of L1451-mm in Feature~B: Leaf 66 is the
strongest structure in the \HCO{} dendrogram, with a peak intensity of
2.78~\Jybm, leaf 30 is the strongest structure in the HCN dendrogram,
with a peak intensity of 2.58~\Jybm, and leaf 10 is the strongest
structure in the \NtwoH{} dendrogram, with a peak intensity of
1.62~\Jybm.

\subsection{Dendrogram Spatial and Kinematics Properties}
\label{sec:ch3sec6sub2}   
The leaves and branches of each dendrogram represent molecular
structures. We fit for the spatial properties of each structure, as we
did in Paper~I, using the {\tt regionprops}
program in MATLAB. This program fits an ellipse to the integrated
intensity footprint of each dendrogram structure to determine its RA
centroid, DEC centroid, major axis, minor axis, and position angle.
Columns~2--6 in Tables~\ref{tbl:dendrohcop}--\ref{tbl:dendron2hp} list
the spatial properties of each structure. To quantify the shape of
each structure, we use the axis ratio and filling factor of the fit
(Columns 7 and 8 of
Tables~\ref{tbl:dendrohcop}--\ref{tbl:dendron2hp}). We lastly define
the structure size (Column 9 of
Tables~\ref{tbl:dendrohcop}--\ref{tbl:dendron2hp}) as the geometric
mean of the major and minor axes, assuming a distance of 235~pc when
converting to parsec units.

\begin{figure}[!h]
\centering
\includegraphics[scale=0.6]{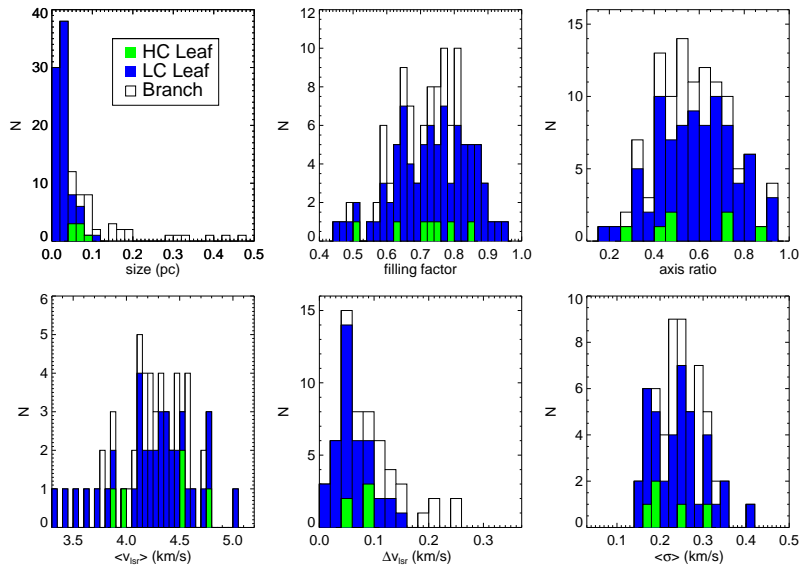}
\caption{ \small Histograms of \HCO{} dendrogram leaf and branch
  properties. High-contrast (HC) leaves, above 6-$\sigma_{n}$
  contrast, are represented by green; low-contrast (LC) leaves, below
  6-$\sigma_{n}$ contrast, are represented by blue; branches are
  represented by white. See the text in Section~\ref{sec:ch3sec6sub2}
  for a discussion of trends seen in these histograms.}
\label{fig:histodendrohcop}
\end{figure}

\begin{figure}[!h]
\centering
\includegraphics[scale=0.6]{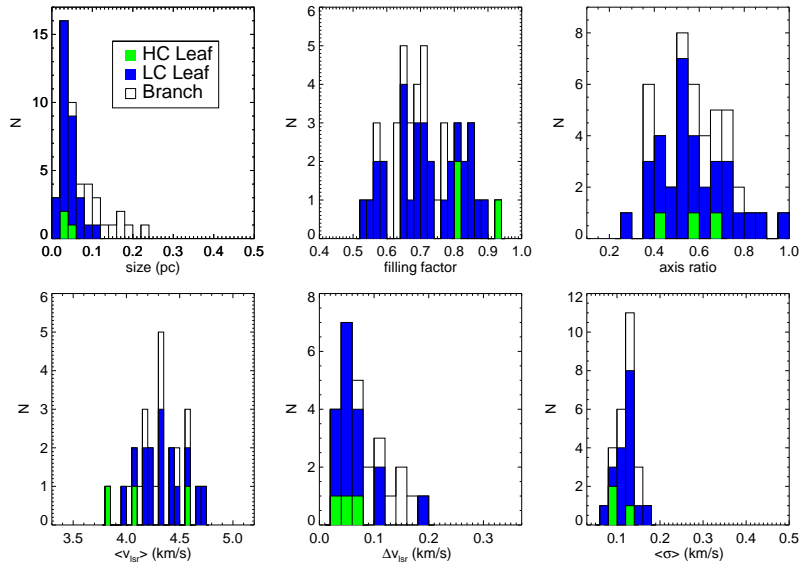}
\caption{ \small Same as Figure~\ref{fig:histodendrohcop}, but for
  HCN.}
\label{fig:histodendrohcn}
\end{figure}

\begin{figure}[!h]
\centering
\includegraphics[scale=0.6]{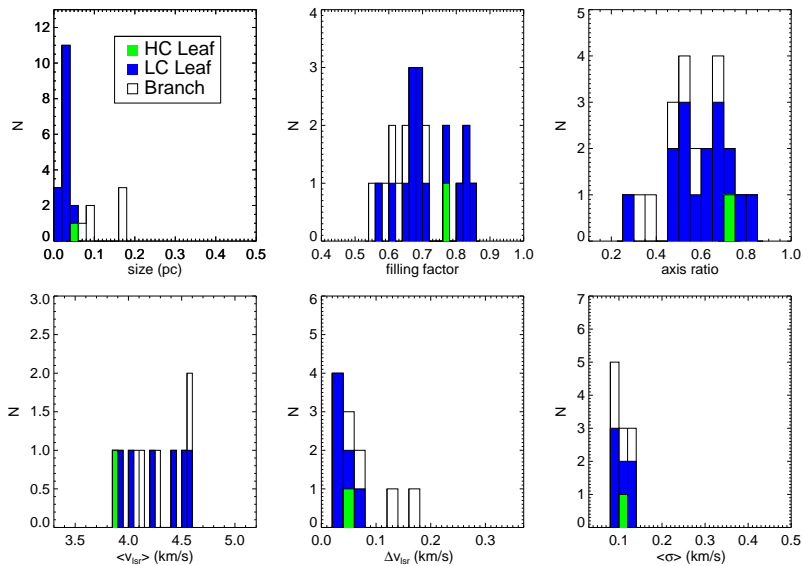}
\caption{ \small Same as Figure~\ref{fig:histodendrohcop}, but for
  \NtwoH{}.}
\label{fig:histodendron2hp}
\end{figure}

Histograms of the size, filling factor, and axis ratio for all leaves
and branches are plotted in the top rows of
Figures~\ref{fig:histodendrohcop}--\ref{fig:histodendron2hp}. All
\HCO{} and HCN leaves are smaller than 0.12~pc, while all \NtwoH{}
leaves are smaller than 0.06~pc. \HCO{} branches are the largest of
all molecules, peaking at 0.46~pc, while HCN branches peak at 0.24~pc,
and \NtwoH{} branches peak at 0.17~pc. This is expected since \HCO{}
shows the most widespread emission in the channel and integrated
emission maps. The filling factor for all molecular structures is
between 0.45 and 0.95. The structures with filling factor closest to
unity are leaves, indicating that leaves are more likely to be
elliptically shaped objects while branches are more likely to be
irregularly shaped objects. The axis ratio for all structures is
between 0.19 and 0.95, showing that there is a distribution from
elongated to round structures, without strong differences between
leaves and branches. There are no obvious differences between the
high-contrast leaves and the rest of the leaves.

We use the integrated intensity footprints of the dendrogram
structures in combination with the centroid velocity and velocity
dispersion maps to determine kinematic properties of leaves and
branches. The four properties present in
Tables~\ref{tbl:dendrohcop}--\ref{tbl:dendron2hp} are: mean and rms
centroid velocity ($\langle V_{\mathrm{lsr}} \rangle$ and $\Delta
V_{\mathrm{lsr}}$, respectively), and mean and rms velocity dispersion
($\langle \sigma \rangle$ and $\Delta \sigma$, respectively). We
illustrated how to derive these properties for leaves and branches in
Section~6 of Paper~I.

Histograms of $\langle V_{\mathrm{lsr}} \rangle$, $\Delta
V_{\mathrm{lsr}}$, and $\langle \sigma \rangle$ are plotted in the
bottom rows of
Figures~\ref{fig:histodendrohcop}--\ref{fig:histodendron2hp}. \HCO{}
traces larger variation in $\langle V_{\mathrm{lsr}} \rangle$ compared
to HCN and \NtwoH{}. We attribute this to \HCO{} being sensitive to
more widespread emission away from the densest regions of L1451, and
therefore tracing more widespread centroid velocities relative to the
systemic velocity of the most spatially compact gas. For all
molecules, $\Delta V_{\mathrm{lsr}}$ of the leaves generally extends
from low to moderate velocities, while it is distributed from moderate
to high velocities for the branches. This indicates a trend where
$\Delta V_{\mathrm{lsr}}$ is generally lower for smaller structures
than larger structures. This trend was also seen for Barnard~1 gas
structures, and we discussed explanations in Paper~I; the primary
reason was the scale-dependent nature of turbulence, which causes gas
parcels separated by smaller distances to have smaller rms velocity
differences between them \citep{2007ARAA..45..565M}.

The distribution of $\langle \sigma \rangle$ is similar for leaves and
branches, with neither distribution showing a preference to peak at
high or low velocities. This trend was also seen in Paper~I for
Barnard~1, indicating that $\langle \sigma \rangle$ does not strongly
depend on the projected size of a structure. The peak $\langle \sigma
\rangle$ for \HCO{} is higher compared to HCN and \NtwoH{}: 0.42~\kms,
0.18, and 0.13~\kms, respectively. Since \HCO{} traces effective
excitation densities of an order of magnitude lower than HCN and
\NtwoH{} \citep{2015PASP..127..299S}, we are likely observing emission
from lower-density gas that is more extended along the
line-of-sight. This could increase the line-of-sight velocity
dispersions, which are expected to scale with cloud depth in a
turbulent medium \citep{2007ARAA..45..565M}.  There are no obvious
differences in these kinematic properties between the high-contrast
leaves and the rest of the leaves.

\subsection{Tree Statistics}
\label{sec:ch3sec6sub3}   
The dendrograms in
Figures~\ref{fig:hcopdendrogram}--\ref{fig:n2hpdendrogram} show an
apparently wide variety of hierarchical complexity between molecular
tracers and between sub-regions of L1451.  In this section, we
quantify the hierarchical nature of the gas with tree statistics that
were introduced and explained in Paper~I so that the complexity
between molecules and sub-regions can be quantitatively
compared. Specifically, we calculate the maximum branching level, mean
path length, and mean branching ratio of the entire L1451 region for
each molecule in a uniform way that accounts for differences in the
noise-level of each data cube (see Appendix~\ref{app:appendixC}).  We
then calculate those same statistics for individual features within
each dendrogram.

To compare the tree statistics of the dendrograms from the different
molecular tracers we follow Appendix~\ref{app:appendixC} and only
consider leaves and branches above a 2.5-$\sigma_{n}$ intensity cut,
where $\sigma_{n}$ is the rms of the noisiest molecular data cube. The
\NtwoH{} PPV cube has the highest noise level, at $\sim$0.14~\Jybm{},
so the cut for each dendrogram is at $\sim$0.35~\Jybm{}, and is
represented as the horizontal dotted line in
Figures~\ref{fig:hcopdendrogram}--\ref{fig:n2hpdendrogram}.  Only
leaves that peak at least 2-$\sigma_{n}$ above the cut are still
considered in the statistics---all other leaves are marked with an
``x'' in the dendrograms. A branch below the cut is discounted, but if
the leaf directly above it is more than 2-$\sigma_{n}$ above the cut,
then the leaf is counted as a leaf with a branching level of zero
(e.g., leaf 64 in Figure~\ref{fig:hcopdendrogram}). This comparison of
tree statistics ensures that we can compare the hierarchical structure
of emission from the different molecules independent of noise-level
differences introduced from the observing setup. These considerations
are not needed when comparing tree statistics from sub-regions of a
single dendrogram.

The path length statistic, defined only for leaves, is the number of
branching levels it takes to go from a leaf to the tree base. The
branching ratio statistic, defined only for branches, is the number of
structures a branch splits into immediately above it in the
dendrogram.  We will be linking these tree statistics to cloud
fragmentation in the discussion to follow. A more evolved region with
a lot of hierarchical structure will have a higher maximum branching
level and larger mean path length than a young region that is starting
to form overdensities and fragment.  The branching ratio of a very
hierarchical region will be smaller than a region fragmenting into
many substructures in a single step.  This is likely an overly
simplistic view, since the molecular emission, and in turn, the
dendrograms and tree statistics, can also be affected by projection
effects and line opacity. We briefly discussed these effects in
Paper~I; since they are extremely difficult to accurately model over
such a large area, we use the simple view that hierarchy comes from
fragmentation\footnote{Although a few regions within L1451 show
  double-peaked \HCO{} spectra, which we attribute primarily to
  self-absorption and not true multiple components, the dendrogram
  algorithm rarely splits structures containing such spectral features
  in two. We searched the dendrogram structure cube and data cube by
  eye, and determined that only leaves 15 and 47 are likely split due
  to these double-peaked spectra. Accounting for this would reduce the
  branching ratio of branch 93 from three to two, have no effect on
  the maximum branching level, and have negligible impact on the
  average tree statistics. Therefore, we do not consider the \HCO{}
  dendrogram to be contaminated by double-peaked spectra.}.

The tree statistics of each dendrogram from the different molecular
tracers are reported in the first section of
Table~\ref{tbl:treestats}. The mean branching ratios are 3.8, 3.3, and
3.1 for the \HCO{}, HCN, and \NtwoH{} emission, respectively. We
interpret this to mean that each molecule is tracing physical
structures that are fragmenting in a hierarchically similar way (e.g.,
a structure is most likely to fragment into about three to four
sub-structures). The mean path length of \HCO{} is 1.0 level longer
than that of HCN and \NtwoH{}, and the maximum branching level of
\HCO{} is two more than that of HCN and three more than that of
\NtwoH{}.  This trend in fragmentation levels is likely due to the
ability of each tracer to detect material at different spatial scales
and physical densities.  As the effective excitation density of the
tracer goes down from \NtwoH{} to HCN to \HCO{}, our observations are
more sensitive to more widespread emission, which means we are
sensitive to more levels of fragmentation extending from the
higher-density leaves (detectable with all tracers) to the lowest
detectable branches (most detectable with \HCO{}). Therefore, even
though the dendrograms in
Figures~\ref{fig:hcopdendrogram}--\ref{fig:n2hpdendrogram} look very
different, a comparison of their tree statistics using a uniform
noise-level cut, along with an understanding of what each molecule is
tracing, produces a consistent picture of the hierarchical structure
of dense gas in L1451 from $\sim$0.5~pc to sub-0.1~pc scales.

The tree statistics of sub-regions from individual molecular tracers
are reported in the second section of Table~\ref{tbl:treestats}. We
compare the sub-region statistics of Features A, B, C, and all other
features with hierarchical complexity (e.g., Feature H in the \HCO{}
and HCN dendrograms). The sections of the dendrograms corresponding to
individual features are marked in
Figures~\ref{fig:hcopdendrogram}--\ref{fig:n2hpdendrogram} with letter
identifiers, and we only consider structures above those identifiers
in this comparison. For example, in the \HCO{} dendrogram, Features A
and C merge together at the branch labeled ``A+C,'' but we consider
the statistics of the individual features only above labels ``A'' and
``C''.  Features A, B, and C are the only features with any
hierarchical complexity in the \NtwoH{} dendrogram, and are the ones
with the most branching steps in the HCN and \HCO{} dendrograms. They
are also the most likely sites for current and future star formation,
since they account for emission surrounding the only continuum source
detections in the field: Feature B surrounds L1451-mm, and Features A
and C surround Bolocam 1~mm sources.

There is a trend of decreasing maximum branching level and mean path
length from Feature B to A to C and then to the other remaining
features.  Features A and B are more similar to each other than either
is to the remaining other features, indicating that the gas in both
features has fragmented a similar amount relative to the rest of the
complex structure in the L1451 field. The mean path length and maximum
branching level of Feature C bridge the gap between the maximum
fragmentation amount seen in Feature A and B and the minimum
fragmentation amount seen in the remaining other features.

We interpret the similarity in hierarchical branching levels between
Features A and B to mean that these sub-regions have progressed to a
similar stage along the evolutionary track of cloud fragmentation.  We
know a young star or first core (L1451-mm) is forming within Feature~B
at or near the location of the maximum gas emission intensity (leaf
66, 30, and 10 for \HCO{}, HCN, and \NtwoH{}, respectively). For
Feature A, PerBolo~6 is at or near the location of maximum gas
emission intensity in Feature A (leaf 15, 7, and 6 for \HCO{}, HCN,
and \NtwoH{}, respectively).  We argue that with Feature A and B
showing very similar tree statistics, with Feature B having a
confirmed compact continuum detection at its hierarchical peak, and
with Feature~A having a single-dish continuum detection at its
hierarchical peak, that a star is likely to form within
Feature~A. This argument can be extended to Feature~C being the next
most likely place for current or future star formation, followed by
the even less fragmented features. Follow-up observations of the
single-dish cores and other column density enhancements in these
features will be useful for testing this expectation. The mean
branching ratios between all features are similar for all molecules,
indicating that all structures are fragmenting into a similar number
of sub-structures at each branching step, regardless how far a feature
is along its evolution toward forming stars.

\begin{deluxetable}{l l l c c c c c c c c c c c c c}
\tabletypesize{\tiny}
\tablecaption{\HCO{} Dendrogram Leaf and Branch Properties}
\tablewidth{0pt}
\setlength{\tabcolsep}{0.03in}
\tablehead
{
 \colhead{{No.}} & \colhead{{RA}} & \colhead{{DEC}} & \colhead{{Maj. Axis}} & \colhead{{Min. Axis}} & \colhead{{PA}} & \colhead{{Axis}} & \colhead{{Filling}} & \colhead{{Size}} & \colhead{{$\langle V_{\mathrm{lsr}} \rangle$}} & \colhead{{$\Delta V_{\mathrm{lsr}}$}} & \colhead{{$\langle$ $\sigma$ $\rangle$}} & \colhead{{$\Delta \sigma$}}  & \colhead{{Pk. Int.}}  & \colhead{{Contrast}}  & \colhead{{Level}}\\
 \colhead{{}} & \colhead{{(h:m:s)}} & \colhead{{(\degree:\arcmin:\arcsec)}} & \colhead{{(\arcsec)}} & \colhead{{(\arcsec)}} & \colhead{{(\degree)}} & \colhead{{Ratio}} & \colhead{{Factor}} & \colhead{{(pc)}} & \colhead{{(\kms)}} & \colhead{{(\kms)}} & \colhead{{(\kms)}} & \colhead{{(\kms)}} & \colhead{{(\Jybm)}} & \colhead{{}} & \colhead{{}}\\
 \colhead{{(1)}} & \colhead{{(2)}} & \colhead{{(3)}} & \colhead{{(4)}} & \colhead{{(5)}} & \colhead{{(6)}} & \colhead{{(7)}} & \colhead{{(8)}} & \colhead{{(9)}} & \colhead{{(10)}} & \colhead{{(11)}} & \colhead{{(12)}} & \colhead{{(13)}} & \colhead{{(14)}} & \colhead{{(15)}} & \colhead{{(16)}}

}
\startdata
\multicolumn{16}{c}{\bf{Leaves}}\\
\tableline
\noalign{\smallskip}
           13   &   03:25:31.6   &   +30:21:34.6   &          18.1   &          14.7   &         105.5   &          0.81   &          0.86   &         0.019   &          4.72(2)   &          0.04(2)   &          0.21(2)   &          0.04(2)   &          1.29   &           2.2    &    4   \\
            14   &   03:25:01.2   &   +30:21:13.7   &         140.5   &          65.4   &         139.4   &          0.47   &          0.62   &         0.109   &          4.63(2)   &          0.09(2)   &          0.21(2)   &          0.07(1)   &          0.99   &           4.8    &    0   \\
            15   &   03:25:26.1   &   +30:21:54.4   &          56.7   &          42.0   &         134.6   &          0.74   &          0.71   &         0.056   &          4.55(1)   &          0.05(1)   &          0.30(1)   &          0.05(1)   &          2.12   &           6.0    &    5   \\
            16   &   03:25:31.1   &   +30:22:56.5   &         117.4   &          53.9   &         121.4   &          0.46   &          0.75   &         0.091   &          4.76(1)   &          0.06(0)   &          0.17(0)   &          0.04(0)   &          1.74   &           7.4    &    2   \\
            17   &   03:24:55.8   &   +30:24:31.8   &          22.8   &          13.1   &          81.8   &          0.57   &          0.86   &         0.020   &          4.46(3)   &          0.06(2)   &          0.41(1)   &          0.03(1)   &          1.09   &           3.3    &    3   \\
\noalign{\smallskip}
\tableline
\noalign{\smallskip}
\multicolumn{16}{c}{\bf{Branches}}\\
\tableline
\noalign{\smallskip} 
                86   &   03:25:09.8   &   +30:23:52.9   &          47.6   &          42.9   &         118.3   &          0.90   &          0.73   &         0.051   &          3.89(1)   &          0.06(1)   &          0.19(1)   &          0.03(0)   &          2.35   &           ...    &    6   \\
            87   &   03:25:03.4   &   +30:24:39.6   &          50.3   &          38.0   &          54.4   &          0.75   &          0.65   &         0.050   &          4.32(2)   &          0.07(1)   &          0.29(1)   &          0.04(1)   &          1.91   &           ...   &    6   \\
            88   &   03:25:02.7   &   +30:24:17.9   &         110.8   &          59.8   &          49.9   &          0.54   &          0.68   &         0.093   &          4.20(2)   &          0.10(1)   &          0.26(1)   &          0.05(1)   &          1.77   &           ...    &    5   \\
            89   &   03:25:09.3   &   +30:24:03.8   &          89.4   &          53.3   &         161.4   &          0.60   &          0.65   &         0.079   &          4.03(2)   &          0.11(2)   &          0.21(1)   &          0.03(1)   &          1.77   &           ...    &    5   \\
            90   &   03:25:05.0   &   +30:24:13.9   &         186.3   &         102.2   &          93.1   &          0.55   &          0.75   &         0.157   &          4.18(1)   &          0.16(1)   &          0.24(0)   &          0.04(0)   &          1.62   &           ...   &    4   \\
\enddata
\vspace{-0.5cm} \footnotesize
\tablecomments{Table~\ref{tbl:dendrohcop} is published in its entirety
  in a machine readable format in the \textit{Astrophysical Journal}
  online edition of this paper. A portion is shown here for guidance
  regarding its form and content.\\} \tablecomments{(2)--(6) The
  position, major axis, minor axis, and position angle were determined
  from {\tt regionprops} in MATLAB. We do not report formal
  uncertainties of these values since the spatial properties of
  irregularly shaped objects is dependent on the chosen method.
  \\ (7) Axis ratio, defined as the ratio of the minor axis to the
  major axis.\\ (8) Filling factor, defined as the area of the leaf or
  branch inscribed within the fitted ellipse, divided by the area of
  the fitted ellipse. \\ (9) Size, defined as the geometric mean of
  the major and minor axes, for an assumed distance of 235~pc. \\ (10)
  The weighted mean $V_{\mathrm{lsr}}$ of all fitted values within a
  leaf or branch. Weights are determined from the statistical
  uncertainties reported by the IDL MPFIT program. The error in the
  mean is reported in parentheses as the uncertainty in the last
  digit. It was computed as the standard error of the mean, $\Delta
  V_{\mathrm{lsr}}/\sqrt N$, where $\Delta V_{\mathrm{lsr}}$ is the
  value in column 11 and $N$ is the number of beams' worth of pixels
  within a given object.  We report kinematic properties only for
  objects that have at least three synthesized beams' worth of
  kinematic pixels.  \\ (11) The weighted standard deviation of all
  fitted $V_{\mathrm{lsr}}$ values within a leaf or branch. The error
  was computed as the standard error of the standard deviation,
  $\Delta V_{\mathrm{lsr}}/\sqrt{2(N-1)}$, assuming the sample of
  beams was drawn from a larger sample with a Gaussian velocity
  distribution. \\ (12) The weighted mean velocity dispersion of all
  fitted values within a leaf or branch. The error was computed as the
  standard error of the mean, $\Delta\sigma/\sqrt N$.  \\ (13) The
  weighted standard deviation of all fitted velocity dispersion values
  within a leaf or branch. The error was computed as the standard
  error of the standard deviation, $\Delta\sigma/\sqrt{2(N-1)}$.
  \\ (14) For a leaf, this is the peak intensity measured in a single
  channel in the dendrogram analysis. For a branch, this is the
  intensity level where the leaves above it merge together.  \\ (15)
  ``Contrast'' is defined as the difference between the peak intensity
  of a leaf and the height of its closest branch in the dendrogram,
  divided by 1-$\sigma_{n}$. \\ (16) The branching level in the
  dendrogram. For example, the base of the tree is level 0, so an
  isolated leaf that grows directly from the base is considered to be
  at level 0. A leaf that grows from a branch one level above the base
  will be at level 1, etc. \\}
\label{tbl:dendrohcop}
\end{deluxetable}

\clearpage

\begin{deluxetable}{l l l c c c c c c c c c c c c c}
\tabletypesize{\tiny}
\tablecaption{HCN Dendrogram Leaf and Branch Properties}
\tablewidth{0pt}
\setlength{\tabcolsep}{0.03in}
\tablehead
{
 \colhead{{No.}} & \colhead{{RA}} & \colhead{{DEC}} & \colhead{{Maj. Axis}} & \colhead{{Min. Axis}} & \colhead{{PA}} & \colhead{{Axis}} & \colhead{{Filling}} & \colhead{{Size}} & \colhead{{$\langle V_{\mathrm{lsr}} \rangle$}} & \colhead{{$\Delta V_{\mathrm{lsr}}$}} & \colhead{{$\langle$ $\sigma$ $\rangle$}} & \colhead{{$\Delta \sigma$}}  & \colhead{{Pk. Int.}}  & \colhead{{Contrast}}  & \colhead{{Level}}\\
 \colhead{{}} & \colhead{{(h:m:s)}} & \colhead{{(\degree:\arcmin:\arcsec)}} & \colhead{{(\arcsec)}} & \colhead{{(\arcsec)}} & \colhead{{(\degree)}} & \colhead{{Ratio}} & \colhead{{Factor}} & \colhead{{(pc)}} & \colhead{{(\kms)}} & \colhead{{(\kms)}} & \colhead{{(\kms)}} & \colhead{{(\kms)}} & \colhead{{(\Jybm)}} & \colhead{{}} & \colhead{{}}\\
 \colhead{{(1)}} & \colhead{{(2)}} & \colhead{{(3)}} & \colhead{{(4)}} & \colhead{{(5)}} & \colhead{{(6)}} & \colhead{{(7)}} & \colhead{{(8)}} & \colhead{{(9)}} & \colhead{{(10)}} & \colhead{{(11)}} & \colhead{{(12)}} & \colhead{{(13)}} & \colhead{{(14)}} & \colhead{{(15)}} & \colhead{{(16)}}

}
\startdata
\multicolumn{16}{c}{\bf{Leaves}}\\
\tableline
\noalign{\smallskip}
             5   &   03:25:24.1   &   +30:21:03.2   &          71.4   &          41.7   &         162.3   &          0.58   &          0.59   &         0.062   &          4.49(1)   &          0.05(1)   &          0.13(1)   &          0.03(0)   &          1.29   &           3.5    &    4   \\
             6   &   03:25:25.4   &   +30:21:45.4   &          27.1   &          19.7   &          82.7   &          0.73   &          0.86   &         0.026   &          4.56(2)   &          0.06(1)   &          0.15(1)   &          0.04(1)   &          1.30   &           3.6    &    4   \\
             7   &   03:25:26.3   &   +30:22:10.6   &          37.0   &          20.7   &         112.3   &          0.56   &          0.80   &         0.031   &          4.60(2)   &          0.06(1)   &          0.13(1)   &          0.06(1)   &          1.51   &           6.0    &    3   \\
             8   &   03:25:31.6   &   +30:22:59.0   &          53.1   &          29.1   &         101.8   &          0.55   &          0.72   &         0.045   &          4.73(1)   &          0.04(1)   &          0.11(1)   &          0.04(1)   &          0.89   &           2.7    &    2   \\
             9   &   03:25:28.4   &   +30:23:09.9   &          46.9   &          26.4   &          18.9   &          0.56   &          0.57   &         0.040   &       ...   &       ...   &       ...   &       ...   &          0.68   &           2.3    &    1   \\
\noalign{\smallskip}
\tableline
\noalign{\smallskip}
\multicolumn{16}{c}{\bf{Branches}}\\
\tableline
\noalign{\smallskip} 
            33   &   03:25:03.0   &   +30:24:41.8   &          58.4   &          46.1   &          37.9   &          0.79   &          0.67   &         0.059   &          4.30(2)   &          0.08(2)   &          0.15(1)   &          0.05(1)   &          1.14   &           ...    &    3   \\
            34   &   03:25:09.5   &   +30:24:03.4   &          92.4   &          47.2   &         158.3   &          0.51   &          0.78   &         0.075   &          4.00(2)   &          0.10(1)   &          0.11(0)   &          0.03(0)   &          1.14   &           ...    &    2   \\
            35   &   03:25:02.7   &   +30:24:32.7   &         112.1   &          44.4   &          37.4   &          0.40   &          0.68   &         0.080   &          4.28(2)   &          0.11(2)   &          0.15(1)   &          0.05(1)   &          0.99   &           ...    &    2   \\
            36   &   03:25:05.6   &   +30:24:16.7   &         180.1   &         100.9   &         103.4   &          0.56   &          0.75   &         0.154   &          4.17(1)   &          0.14(1)   &          0.12(1)   &          0.06(0)   &          0.85   &           ...    &    1   \\
            37   &   03:25:24.6   &   +30:21:11.0   &          98.8   &          64.4   &          21.4   &          0.65   &          0.65   &         0.091   &          4.50(1)   &          0.06(1)   &          0.14(1)   &          0.04(0)   &          0.78   &           ...    &    3   \\
\enddata
\vspace{-0.5cm} \footnotesize
\tablecomments{Table~\ref{tbl:dendrohcn} is published in its
    entirety in a machine readable format in the \textit{Astrophysical
      Journal} online edition of this paper. A portion is shown here
    for guidance regarding its form and content.\\}
\tablecomments{Columns the same at Table~\ref{tbl:dendrohcop}.}
\label{tbl:dendrohcn}
\end{deluxetable}

\clearpage

\begin{deluxetable}{l l l c c c c c c c c c c c c c}
\tabletypesize{\tiny}
\tablecaption{\NtwoH{} Dendrogram Leaf and Branch Properties}
\tablewidth{0pt}
\setlength{\tabcolsep}{0.03in}
\tablehead
{
 \colhead{{No.}} & \colhead{{RA}} & \colhead{{DEC}} & \colhead{{Maj. Axis}} & \colhead{{Min. Axis}} & \colhead{{PA}} & \colhead{{Axis}} & \colhead{{Filling}} & \colhead{{Size}} & \colhead{{$\langle V_{\mathrm{lsr}} \rangle$}} & \colhead{{$\Delta V_{\mathrm{lsr}}$}} & \colhead{{$\langle$ $\sigma$ $\rangle$}} & \colhead{{$\Delta \sigma$}}  & \colhead{{Pk. Int.}}  & \colhead{{Contrast}}  & \colhead{{Level}}\\
 \colhead{{}} & \colhead{{(h:m:s)}} & \colhead{{(\degree:\arcmin:\arcsec)}} & \colhead{{(\arcsec)}} & \colhead{{(\arcsec)}} & \colhead{{(\degree)}} & \colhead{{Ratio}} & \colhead{{Factor}} & \colhead{{(pc)}} & \colhead{{(\kms)}} & \colhead{{(\kms)}} & \colhead{{(\kms)}} & \colhead{{(\kms)}} & \colhead{{(\Jybm)}} & \colhead{{}} & \colhead{{}}\\
 \colhead{{(1)}} & \colhead{{(2)}} & \colhead{{(3)}} & \colhead{{(4)}} & \colhead{{(5)}} & \colhead{{(6)}} & \colhead{{(7)}} & \colhead{{(8)}} & \colhead{{(9)}} & \colhead{{(10)}} & \colhead{{(11)}} & \colhead{{(12)}} & \colhead{{(13)}} & \colhead{{(14)}} & \colhead{{(15)}} & \colhead{{(16)}}

}
\startdata
\multicolumn{16}{c}{\bf{Leaves}}\\
\tableline
\noalign{\smallskip}
             6   &   03:25:26.0   &   +30:21:44.0   &          51.1   &          36.0   &          30.7   &          0.70   &          0.66   &         0.049   &          4.57(1)   &          0.04(1)   &          0.13(1)   &          0.03(1)   &          1.25   &           4.0    &    3   \\
             7   &   03:25:03.9   &   +30:25:00.3   &          36.8   &          19.6   &          89.7   &          0.53   &          0.68   &         0.031   &          4.45(1)   &          0.03(1)   &          0.09(1)   &          0.02(0)   &          1.40   &           3.5    &    1   \\
             8   &   03:25:16.1   &   +30:19:47.4   &          35.7   &          28.9   &          54.8   &          0.81   &          0.68   &         0.037   &       ...   &       ...   &       ...   &       ...   &          0.89   &           3.7    &    0   \\
             9   &   03:24:55.8   &   +30:23:23.3   &          19.3   &          11.6   &          61.8   &          0.60   &          0.77   &         0.017   &       ...   &       ...   &       ...   &       ...   &          0.66   &           2.1    &    0   \\
             10   &   03:25:07.3   &   +30:24:32.1   &          38.5   &          21.1   &          96.6   &          0.55   &          0.71   &         0.032   &          4.24(1)   &          0.03(1)   &          0.08(0)   &          0.01(0)   &          1.62   &           3.0    &    2   \\
\noalign{\smallskip}
\tableline
\noalign{\smallskip}
\multicolumn{16}{c}{\bf{Branches}}\\
\tableline
\noalign{\smallskip} 
            16   &   03:25:08.5   &   +30:24:11.8   &         101.3   &          38.3   &         136.7   &          0.38   &          0.71   &         0.071   &          4.11(2)   &          0.12(1)   &          0.10(0)   &          0.02(0)   &          1.18   &           ...    &    1   \\
            17   &   03:25:04.8   &   +30:24:18.4   &         196.1   &         108.9   &          95.0   &          0.56   &          0.62   &         0.166   &          4.26(2)   &          0.17(1)   &          0.11(0)   &          0.03(0)   &          0.90   &           ...    &    0   \\
            18   &   03:25:17.7   &   +30:18:54.8   &         141.7   &          42.6   &         101.8   &          0.30   &          0.65   &         0.088   &          4.08(2)   &          0.07(2)   &          0.13(1)   &          0.04(1)   &          0.69   &          ...    &    0   \\
            19   &   03:25:25.6   &   +30:21:34.2   &          96.5   &          65.4   &          41.1   &          0.68   &          0.62   &         0.091   &          4.58(1)   &          0.04(1)   &          0.10(1)   &          0.02(0)   &          0.67   &           ...    &    2   \\
            20   &   03:25:27.2   &   +30:21:45.9   &         200.6   &         100.1   &          53.0   &          0.50   &          0.56   &         0.161   &       ...   &       ...   &       ...   &       ...   &          0.53   &           ...    &    1   \\
\enddata
\vspace{-0.5cm} \footnotesize
\tablecomments{Table~\ref{tbl:dendron2hp} is published in its
    entirety in a machine readable format in the \textit{Astrophysical
      Journal} online edition of this paper. A portion is shown here
    for guidance regarding its form and content.\\}
\tablecomments{Columns the same as Table~\ref{tbl:dendrohcop}.}
\label{tbl:dendron2hp}
\end{deluxetable}

\clearpage

\begin{deluxetable}{l c c c c}
\tabletypesize{\footnotesize}
\tablecaption{Tree Statistics}
\tablewidth{0pt}
\setlength{\tabcolsep}{0.03in}
\tablehead
{
 \colhead{Line (Sub-region)}  & \colhead{Total No.$^{a}$} & \colhead{Max Level$^{b}$} & \colhead{Mean PL$^{c}$} & \colhead{Mean BR$^{d}$} 
}
\startdata    
\noalign{\smallskip}
\noalign{\smallskip}
\tableline
\noalign{\smallskip}
\multicolumn{5}{c}{\bf{Comparison Across Tracers$^{e}$}}\\
\tableline
\noalign{\smallskip} 
HCO$^{+}$      & 113  & 6 & 2.3 & 3.8 \\
HCN            & 46  & 4 & 1.3 & 3.3 \\
N$_{2}$H$^{+}$ & 22   & 3 & 1.3 & 3.1 \\
\noalign{\smallskip}
\tableline
\noalign{\smallskip}
\multicolumn{5}{c}{\bf{Comparison of Sub-Regions$^{f}$}}\\
\tableline
\noalign{\smallskip}
HCO$^{+}$ (A) & 10  &  6 & 4.8 & 2.3 \\
HCO$^{+}$ (B) & 16  &  7 & 5.9 & 2.5 \\
HCO$^{+}$ (C) & 8  &  5 & 4.2 & 2.3 \\
HCO$^{+}$ (others) & 9  &  4 & 3.7 & 2.0 \\
\hline
HCN (A)       & 13  & 4 & 2.4 & 2.4 \\
HCN (B)       & 13  & 4 & 2.6 & 2.4 \\
HCN (C)       & 6   & 2 & 1.5 & 2.5 \\
HCN (others)  & 3   & 1 & 1.0 & 2.0 \\
\hline
N$_{2}$H$^{+}$ (A) & 9  & 3 & 2.2 & 2.7 \\
N$_{2}$H$^{+}$ (B) & 5 & 2 & 1.7 & 2.0 \\
N$_{2}$H$^{+}$ (C) & 3  & 1 & 1.0 & 2.0 \\
N$_{2}$H$^{+}$ (others) & NA  & NA & NA & NA
\enddata
\vspace{-0.8cm} \tablecomments{ $^{a}$ Total number of leaves and
  branches. $^{b}$ Maximum branching level tree statistic. $^{c}$ Mean
  path length tree statistic. $^{d}$Mean branching ratio tree
  statistic. $^{e}$ Using method for comparing dendrograms of
    data with different noise levels, where only structures above the
    horizontal dotted line in
    Figures~\ref{fig:hcopdendrogram}--\ref{fig:n2hpdendrogram} without
    an ``x'' symbol are used to calculate tree statistics. $^{f}$
    Using original dendrogram of the individual molecular tracer,
    where all structures are used to calculate tree statistics.}
\label{tbl:treestats}
\end{deluxetable}

\clearpage

\section{Dust in L1451}
\label{sec:ch3sec7}

The CLASSy observations provide excellent measurements of gas
structure and kinematics, but are less reliable for column density or
mass information due to large uncertainties in relative abundance and
opacity of the molecular emission.  For this, we turned to
\textit{Herschel} observations.

\subsection{L1451 Column Density and Temperature}
\label{sec:ch3sec7sub1}

We used \textit{Herschel} 160, 250, 350, and 500~$\mu$m observations
of L1451 to derive the column density and temperature of the
dust. There is no detectable emission at 70~$\mu$m toward this region.
The \textit{Herschel} images were corrected for the zero-level offset
based on a comparison with \textit{Planck} and DIRBE/IRAS data
\citep{2015ApJ...798...88M}. In detail, we added offsets of 37.1,
42.6, 23.3 and 10.2 MJy/sr to the \textit{Herschel} 160, 250, 350 and
500~$\mu$m maps, respectively. The images were convolved to the
angular resolution of the \textit{Herschel} 500~$\mu$m band
($\sim$36\arcsec) using the convolution kernels from
\citet{2011PASP..123.1218A}, and were regridded to a common pixel size
of 10\arcsec.  The fitting was performed on a pixel-by-pixel basis
with a modified blackbody spectrum of $I_{\nu} = \kappa_{\nu}B(\nu, T)
\Sigma$, where $\kappa_{\nu}$ is the dust mass opacity coefficient at
frequency $\nu$, $B(\nu, T)$ is the Planck function at temperature T,
and $\Sigma = \mu m_{p} N(H_{2})$ is the gas mass column density for a
mean molecular weight of $\mu=2.8$ \cite[e.g.][]{2008A&A...487..993K}
assuming a gas-to-dust ratio of 100:1.  We assume $\kappa_{\nu}=0.1
\times (\nu / 10^{12} \ Hz)^{\beta}$ cm$^{2}$~g$^{-1}$
\citep{1991ApJ...381..250B} and $\beta=1.7$.

\begin{figure}[!h]
\centering \includegraphics[scale=0.5]{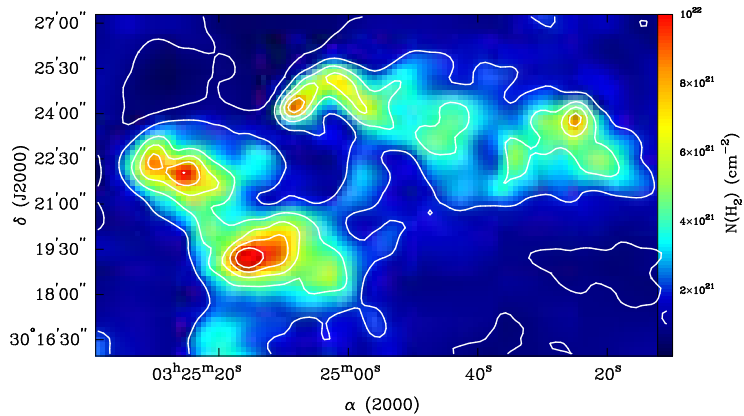}
\caption{ \small Column density map of L1451 derived from
  \textit{Herschel} 160, 250, 350, and 500~$\mu$m data. The white
  contours correspond to N(H$_{2}$)=[0.8, 2.0, 4.0, 6.0, 8.0,
    10.0]$\times$10$^{21}$~cm$^{-2}$. The measured column densities
  toward the densest regions are likely systematically underestimated
  by a factor of two if considering two-temperature models over
  single-temperature models; see the discussion in the text.}
\label{fig:herschelCD}
\end{figure}

\begin{figure}[!h]
\centering
\includegraphics[scale=0.5]{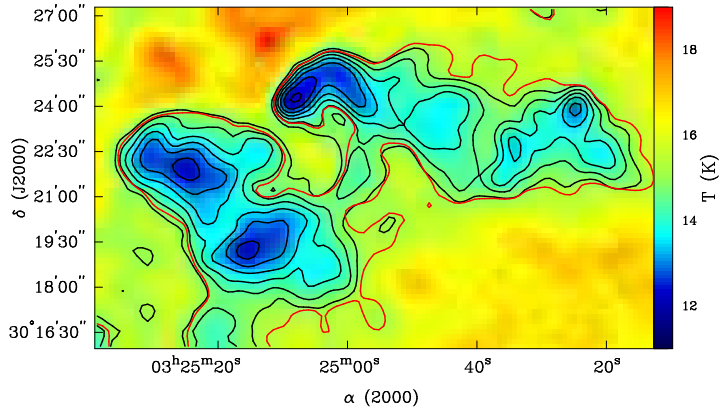}
\caption{ \small Temperature map of L1451 derived from
  \textit{Herschel} 160, 250, 350, and 500~$\mu$m data. The black
  contours represent 12 to 15~K, in 0.5~K increments. The red contour
  represents a column density of 2.0$\times$10$^{21}$~cm$^{-2}$, for
  comparison to Figure~\ref{fig:herschelCD}. The temperatures toward
  the densest regions are likely overestimated by a few Kelvin if
  considering two-temperature models over single-temperature models;
  see text for discussion.}
\label{fig:herschelT}
\end{figure}

The resulting column density and temperature maps are shown in
Figures~\ref{fig:herschelCD} and \ref{fig:herschelT}, respectively. We
validated our SED fitting procedure by comparing our derived optical
depth to that of the \citet{2014A&A...571A..11P} model.  Specifically,
we re-ran our SED fits after smoothing the \textit{Herschel} input
maps to match the \citet{2014A&A...571A..11P} resolution of
5\arcmin{}, and found that our derived 300~$\mu$m optical depth agrees
with that of the \textit{Planck}-based model to within 5\% on average.
The mean column density of L1451 that is enclosed within the
2.0$\times$10$^{21}$~cm$^{-2}$ contour in Figure~\ref{fig:herschelCD}
(the white contour that encircles all of the high column density
regions) is 3.7$\times$10$^{21}$~cm$^{-2}$ with a standard deviation
of 1.7$\times$10$^{21}$~cm$^{-2}$. The peak column density of
1.2$\times$10$^{22}$~cm$^{-2}$ occurs at the location of the Bolocam
source, PerBolo~4.  The mean temperature within the 15.0~K contour of
Figure~\ref{fig:herschelT} is 14.0~K, with a standard deviation of
0.7~K. A minimum temperature of 11.9~K occurs at the location of
L1451-mm.

Gas kinetic temperature measurements exist toward the four Bolocam
sources in our field \citep{2008ApJS..175..509R}.  Those gas
temperatures are $\sim$2--3.5~K lower than we find by fitting the
\textit{Herschel} SEDs. Our results, and those from
\citet{2014A&A...571A..11P} that we compared to, assume emission from
a single cloud layer. However, variations in temperature along the
line of sight, typically caused by a warmer layer of foreground and
background material surrounding a dense, cold star-forming region, are
often present.  This warmer cloud component can drive the fitted
temperature up and fitted column density down when only doing a single
component fit.

To estimate the systematic overestimation of temperatures and
underestimation of column densities toward the densest regions of
clouds, we created a simple radiative transfer model with a cold layer
at 9~K (representing L1451) between two warmer layers at 17~K
(representing foreground and background cloud material). If the warm
layers have $A_{V}$ of 0.4, then the temperature in cold layer regions
with $A_{V} \gtrsim$~2 is overestimated by $\sim$2.5--5.5~K when using
a single component fit (e.g., measure 12~K instead of the true
9~K). This matches the differences between our temperatures and the
temperatures from the ammonia data. Furthermore, measured column
density values of a cold, dense region are predicted to be a factor of
two lower than the true values in regions where the warm layers have
$A_{V}$ of 0.4 and the cold layer has $A_{V} \gtrsim$~2. If the
lower-resolution \textit{Herschel} beam is not filled in regions of
cold, dense gas, then this could further bias temperatures to be too
warm, and column densities to be too low.  In
Section~\ref{sec:ch3sec8}, we will discuss how these systematic
uncertainties can affect our virial analysis results. An upcoming
paper will present a detailed comparison of single-layer and
multi-layer SED fitting across Perseus, quantifying the improvement
that can be achieved for cold, dense regions when considering the hot,
diffuse component along the line-of-sight.

\subsection{Dendrogram Analysis of Dust}
\label{sec:ch3sec7sub2}
The column density results in the previous section are angular
resolution limited compared to our CLASSy maps, so it is difficult to
estimate the mass within the smallest molecular structures we
identified using the dendrogram analysis in Section~\ref{sec:ch3sec6}.
Therefore, we take the approach of first defining structures based on
the dust data here, and then using the kinematic information within
those structures to explore energy balance in the next section.

We converted the N$_{\mathrm{H_{2}}}$ column density map to an
extinction map using a conversion factor of
N$_{\mathrm{H_{2}}}$/$A_{V}$ =
(1/2)~$\times$~1.87~$\times$~10$^{21}$~cm$^{-2}$~mag$^{-1}$
\citep{2003ARA&A..41..241D}. We then ran our non-binary dendrogram
algorithm on the extinction map to define dust structures in the
field. We used an rms and branching step of 0.2~\Av{}, and required
that local maxima peak at least 0.4~\Av{} above the merge level to be
considered a real leaf.  The algorithm identified 8 leaves and 6
branches in the region where we have molecular line data.

\begin{figure}[!h]
\centering \includegraphics[scale=0.63]{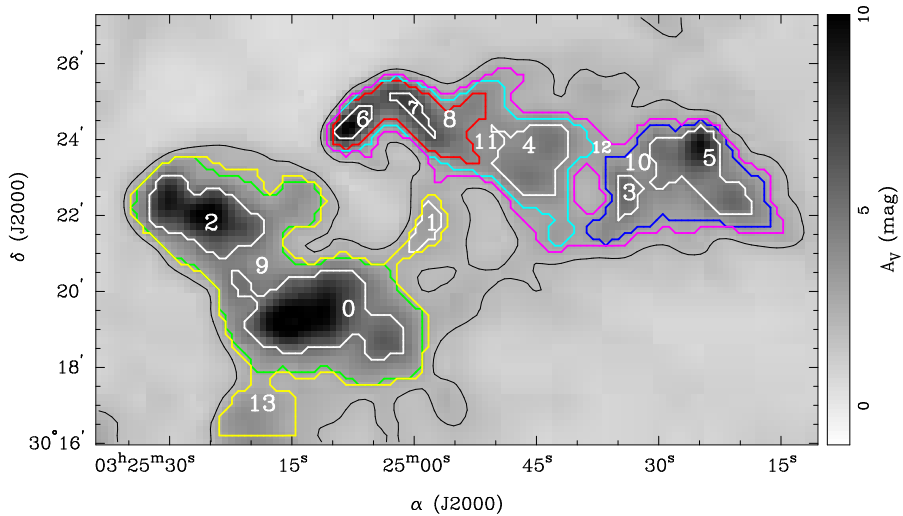}
\caption{ \small Visual extinction map of L1451 (greyscale), as
  derived from the column density map in
  Figure~\ref{fig:herschelCD}. The solid black contour represents
  \Av~=~2, and the mean extinction within that contour is 3.8~mag with
  a 1.8~mag standard deviation. Dendrogram-derived dust structure
  boundaries are shown with colored contours. The peak extinction is
  $\sim$13~mag within structure 0.}
\label{fig:dendrodustlabels}
\end{figure}

The extinction map, with dendrogram-identified structures, is shown in
Figure~\ref{fig:dendrodustlabels}. Properties of the dendrogram
structures are listed in Table~\ref{tbl:dendrodust}, including their
coordinate, size, weighted mean temperature and column density, and
total mass. The mean temperature and column density, and total mass
of each structure considers all of the emission interior to the
structure (e.g., branch 9 includes emission from branch 9 and leaves 0
and 2).  To calculate the total mass within each structure, we first
converted the column density at each pixel to a solar mass unit as:
\begin{equation}
M(\alpha,\delta) = \mu_{H_{2}} m_{H} N_{H_{2}}(\alpha,\delta) A,
\label{eq:ch3eq1}
\end{equation}
where $\mu_{H_{2}}$ is the molecular weight per hydrogen molecule
(2.8), $m_{H}$ is the mass of a hydrogen atom,
$N_{H_{2}}(\alpha,\delta)$ is the column density at a pixel location,
and $A$ is the pixel area. As before, the assumed distance is
235~pc. We then totaled the mass enclosed within each structure. The
mass of leaves ranges from 0.2 to 5.1~\Msun{}, and the mass enclosed
within branches ranges from 3.0 to 15.9~\Msun{}. The mass interior to
the yellow and purple contours in Figure~\ref{fig:dendrodustlabels} is
14.9~\Msun{} and 12.4~\Msun{}, respectively.

\begin{deluxetable}{l l l c c c c c c c }
\tabletypesize{\tiny}
\tablecaption{Dust Structure Properties}
\tablewidth{0pt}
\setlength{\tabcolsep}{0.03in}
\tablehead
{
 \colhead{{No.}} & \colhead{{RA}} & \colhead{{DEC}} & \colhead{{Size}} & \colhead{$\langle$T$\rangle$} & \colhead{$\langle$N$_{\mathrm{H_{2}}}$$\rangle$} & \colhead{M$_{\mathrm{tot}}$} & \colhead{$\sigma_{\mathrm{tot,N_{2}H^{+}}}$} & \colhead{$\sigma_{\mathrm{tot,HCN}}$} & \colhead{$\sigma_{\mathrm{tot,HCO^{+}}}$}  \\
 \colhead{{}} & \colhead{{(h:m:s)}} & \colhead{{(\degree:\arcmin:\arcsec)}} & \colhead{{(pc)}} & \colhead{{(K)}} & \colhead{{(10$^{21}$~cm$^{-2}$)}} & \colhead{{(solar)}} & \colhead{{(\kms)}}  & \colhead{{(\kms)}}  & \colhead{{(\kms)}}  \\
 \colhead{{(1)}} & \colhead{{(2)}} & \colhead{{(3)}} & \colhead{{(4)}} & \colhead{{(5)}} & \colhead{{(6)}} & \colhead{{(7)}} & \colhead{{(8)}} & \colhead{{(9)}} & \colhead{{(10)}}
}
\startdata
\multicolumn{10}{c}{\bf{Leaves}}\\
\tableline
\noalign{\smallskip}
             0   &   03:25:13.3   &   +30:19:05.8   &          0.23   &          13.6   &           5.1   &           5.1   &          0.34   &         0.39 & 0.49    \\
             1   &   03:25:00.2   &   +30:21:22.1   &          0.08   &          14.4   &           2.8   &           0.2   &          ...   &                0.41 & 0.44    \\
             2   &   03:25:27.2   &   +30:21:43.6   &          0.17   &          13.2   &           5.0   &           3.1   &          0.27  &                  0.29 & 0.40    \\
             3   &   03:24:35.5   &   +30:22:09.5   &          0.07   &          13.6   &           4.4   &           0.4   &          0.37   &                0.34 & 0.49    \\
             4   &   03:24:47.0   &   +30:23:11.2   &          0.14   &          13.9   &           4.0   &           1.4   &          ...   &               0.54 & 0.64    \\
             5   &   03:24:26.8   &   +30:22:47.7   &          0.17   &          13.9   &           4.6   &           2.0   &          0.28   &                 0.44 & 0.53    \\
             6$^{a}$   &   03:25:08.7   &   +30:24:09.5   &          0.07   &          12.2   &           7.2   &           0.5   &          0.25   &                  0.25  & 0.33  \\
             7   &   03:25:01.5   &   +30:24:27.9   &          0.07   &          12.7   &           6.4   &           0.5   &          0.26   &              0.28 & 0.40   \\
\noalign{\smallskip}
\tableline
\noalign{\smallskip}
\multicolumn{10}{c}{\bf{Branches}}\\
\tableline
\noalign{\smallskip} 
            8   &   03:25:01.5   &   +30:24:12.9   &          0.21   &          13.7   &           4.1   &           3.0   &          0.33   &                 0.33 & 0.42      \\
            9   &   03:25:17.8   &   +30:20:07.0   &          0.41   &          14.2   &           3.1   &          13.1   &          0.44   &                    0.47  & 0.54  \\
            10   &   03:24:29.2   &   +30:22:26.8   &          0.27   &          14.2   &           3.6   &           4.5   &          0.35   &                 0.45 & 0.57    \\
            11   &   03:24:54.2   &   +30:23:38.9   &          0.33   &          14.1   &           3.3   &           6.2   &          0.34   &                  0.39 & 0.51    \\
            12   &   03:24:43.1   &   +30:23:04.4   &          0.46   &          14.3   &           3.1   &          12.4   &          0.39   &                   0.43 & 0.54  \\
            13   &   03:25:17.1   &   +30:19:52.8   &          0.51   &          14.3   &           3.0   &          14.9   &          0.47   &                  0.50  &  0.56   
\enddata
\vspace{-0.5cm} \tiny \tablecomments{(4) Geometric mean of major and
  minor axis fit to structure (used as diameter of structure). (5)
  Weighted mean temperature within structure. (6) Weighted mean column
  density of H$_{2}$ within structure. (7) Total mass within
  structure. (8) \NtwoH{} velocity dispersion calculated from
  integrated spectrum across structure; see
  Section~\ref{sec:ch3sec8sub2} for discussion. (9) HCN velocity
  dispersion calculated from integrated spectrum across structure; see
  Section~\ref{sec:ch3sec8sub2} for discussion. (10) \HCO{} velocity
  dispersion calculated from integrated spectrum across structure; see
  Section~\ref{sec:ch3sec8sub2} for discussion. $^{a}$ L1451-mm is
  located within this structure.}
\label{tbl:dendrodust}
\end{deluxetable}

\clearpage

\section{Discussion}
\label{sec:ch3sec8}

The goal of this discussion section is to understand the state of star
formation in L1451 using results and analysis presented in previous
sections.  We selected L1451 as a CLASSy region because of its very
low star formation activity. There are no confirmed protostar
detections, and only one confirmed compact continuum core. This is
very different from the other CLASSy regions, which have many
protostars and outflows.  We wanted to use L1451 to study the
structure and kinematics of the densest regions of clouds before star
formation activity feeds back into the cloud.

We now explore the following questions. What column density threshold
are we capturing with our spectral line observations, and do
dendrogram-identified structures trace actual column density features
that can inform structure formation in a young cloud? If so, will any
structures go on to form stars? We address these questions in this
section by exploring the correspondence between molecular line and
continuum emission, with a virial analysis of L1451 structures, and by
describing the similarities and differences between L1451-mm and
L1451-west. We conclude with an analysis of the three-dimensional
morphology of L1451 on the largest scales to see how it compares to
more evolved CLASSy regions.

\subsection{Connecting molecular and column density structures}
\label{sec:ch3sec8sub1}
\begin{figure}[!h]
\centering
\includegraphics[scale=0.35]{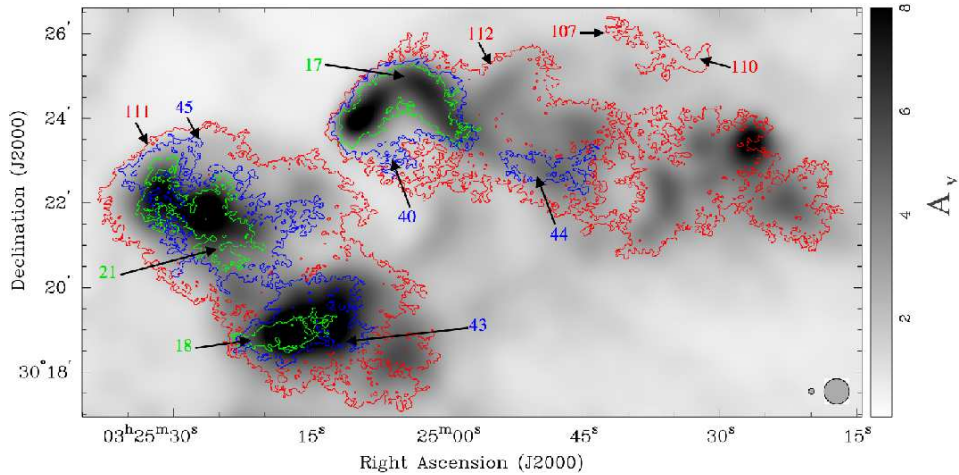}
\caption{\small Dendrogram lowest-level branch footprints for all
  molecules, overplotted on our extinction map that was regridded to
  match the CLASSy pixel scale (beams for column density (36\arcsec{})
  and molecular (8\arcsec{}) maps are in the lower right). Red =
  \HCO{}, blue = HCN, and green = \NtwoH{}. We label each branch with
  its dendrogram structure number, and branch properties can be
  examined in Tables~\ref{tbl:dendrohcop}--\ref{tbl:dendron2hp}.}
\label{fig:dendrobranchmap}
\end{figure}

\begin{figure}[!h]
\centering \includegraphics[scale=0.35]{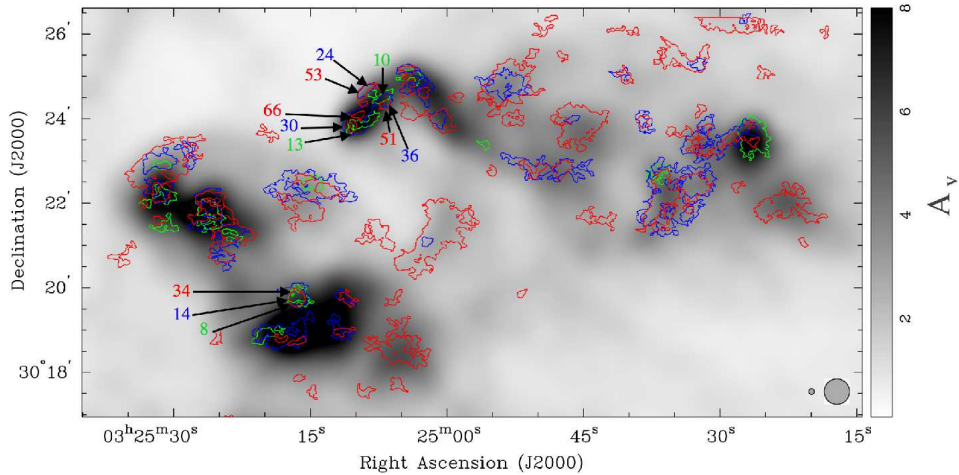}
\caption{\small Dendrogram leaf footprints for all molecules,
  overplotted on our extinction map that was regridded to match the
  CLASSy pixel scale (beams for column density (36\arcsec{}) and
  molecular (8\arcsec{}) maps are in the lower right). Red = \HCO{},
  blue = HCN, and green = \NtwoH{}. We label a few examples with
  dendrogram structure numbers, and leaf properties can be examined in
  Tables~\ref{tbl:dendrohcop}--\ref{tbl:dendron2hp}.}
\label{fig:dendroleafmap}
\end{figure}

L1451 is the one CLASSy region with strong, widespread \HCO{} and HCN
that is not affected by outflows and significant self-absorption. This
enabled a dendrogram analysis of all three molecules in
Section~\ref{sec:ch3sec6}, instead of just \NtwoH{}, and lets us now
compare the identified molecular structures to the column density
structure of L1451 presented in
Section~\ref{sec:ch3sec7}. Figures~\ref{fig:dendrobranchmap} and
\ref{fig:dendroleafmap} have the dendrogram footprints of the
lowest-level branches, and all the leaves, respectively, overplotted
on column density structure derived from \textit{Herschel} data. We
combined the footprints of all the contoured emission in
Figures~\ref{fig:dendrobranchmap} and \ref{fig:dendroleafmap} to
create a mask for the column density map for determining how well the
molecular emission captures material at different column densities.

\begin{figure}[!h]
\centering \includegraphics[scale=0.75]{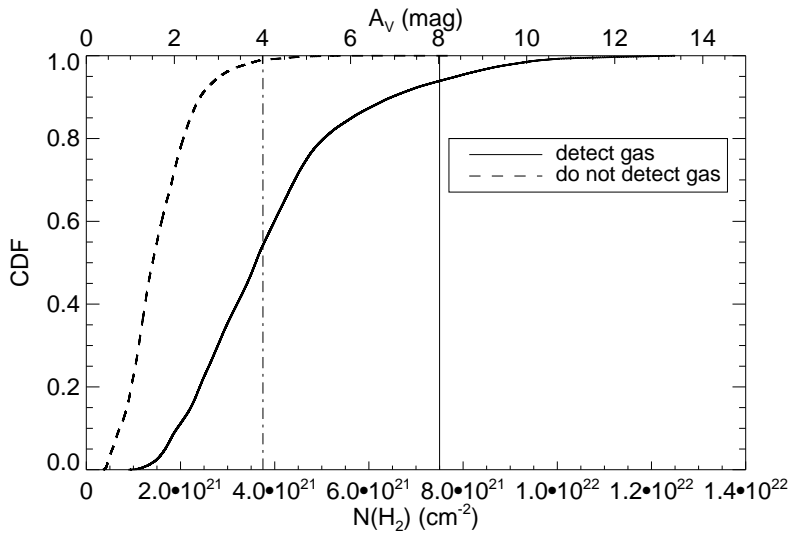}
\caption{ \small Cumulative distribution functions for the areas of
  the L1451 column density map where we have detected molecular
  emission (solid curve), and the areas where we have not detected
  molecular emission (dashed curve). The solid vertical line marks the
  column density threshold for star formation \citep[][and references
    therein]{2014prpl.conf...27A}, while the dashed-dotted vertical
  line represents that threshold if our measured column densities are
  underestimated by a factor of two from the true column densities.}
\label{fig:cdf}
\end{figure}

Figure~\ref{fig:cdf} shows cumulative distribution functions for
column density in regions with and without line emission (regions with
line emission are defined as having at least a single molecular
detection, while regions without line emission are defined as having
no molecular detections). A threshold for star formation above
\Av{}~$\sim$~8, or an H$_{\mathrm{2}}$ column density of
7.5~$\times$~10$^{21}$~cm$^{-2}$, has been postulated based on the
distribution of prestellar cores and protostars within the densest
regions of molecular clouds \citep[][and references
  therein]{2004ApJ...611L..45J,2005A&A...440..151H,2014prpl.conf...27A}.
Only 0.002\% of dust regions without a molecular gas detection are
above the threshold column density if we take our derived column
densities as correct. If our measured column densities are uniformly
underestimated by a factor of two from the true values, as was
discussed in Section~\ref{sec:ch3sec7sub1}, still only 1\% of dust
regions without a molecular gas detection are above the threshold
column density.  90\% of the regions where we detect molecules are at
column densities above 1.9~$\times$~10$^{21}$~cm$^{-2}$, with a
maximum of 1.3~$\times$~10$^{22}$~cm$^{-2}$, and minimum of
8.9~$\times$~10$^{20}$~cm$^{-2}$. 90\% of the regions where we do not
detect molecules are below 2.4~$\times$~10$^{21}$~cm$^{-2}$, with a
maximum of 7.8~$\times$~10$^{21}$~cm$^{-2}$, and minimum of
3.6~$\times$~10$^{20}$~cm$^{-2}$. This shows that spectral line
observations using our suite of dense-gas tracer molecules are a great
probe of the star forming material in young regions above the
threshold for star formation and down to column densities of a
few~$\times$~10$^{21}$~cm$^{-2}$.

This result of molecular emission capturing most of the cloud material
near and above the threshold of star formation, combined with the
result that the branches in Figure~\ref{fig:dendrobranchmap} are
fragmenting to form the leaves in Figure~\ref{fig:dendroleafmap} in a
similar hierarchical fashion for each molecule (see
Section~\ref{sec:ch3sec6sub3}), suggests that dendrogram-identified
molecular structure is able to trace physical structure formation,
despite some biases due to chemistry and extinction.  This provides
observational evidence that structure formation precedes star
formation in molecular clouds. Complex morphological structure in a
turbulent cloud can be formed by turbulence-driven cascades from
large-scale flows before the onset of star formation \citep[][and
  reference therein]{2010A&A...520A..17K}, and it can be produced from
internally-driven turbulence from protostellar feedback
\citep{2009ApJ...695.1376C,2014ApJ...790..128F,2014ApJ...783..115N}.
Our result helps to disentangle externally and internally driven
structure, which it is important for demonstrating an observational
case of complex, hierarchical dense gas structure existing in a
turbulent cloud at an epoch before internal feedback can impact the
natal cloud environment.

\subsection{Virial Analysis of Structures}
\label{sec:ch3sec8sub2}

We next use a virial analysis to assess the stability of L1451
structures against collapse. We know one star is currently forming in
L1451, based on the detection of compact 3~mm emission at
L1451-mm. However, it is unknown if the majority of L1451 structures
are dominated by gravitational potential energy and therefore unstable
against collapse, or rather dominated by internal pressure and
therefore stable against collapse and possibly confined by large
external pressure. For the virial analysis, we use the dust results in
combination with the molecular data. In Section~\ref{sec:ch3sec7}, we
derived column densities and temperatures across L1451 at the angular
resolution of the longest wavelength \textit{Herschel} band
(36\arcsec{}). We also found dendrogram-identified dust structures in
the extinction map.  In this section, we use that information for dust
leaves along with CLASSy kinematic data from Section~\ref{sec:ch3sec5}
to assess the energy balance of several structures in L1451.

The virial theorem is useful for describing the stability of
structures. There are several approaches for applying the virial
theorem to molecular cloud observations in the literature
\citep[e.g.,][]{1981MNRAS.194..809L,1992ApJ...395..140B,1992ApJ...399..551M,2006MNRAS.372..443B,2013ApJ...779..185K}. We
follow the formalism of \citet{1992ApJ...395..140B} and
\citet{2013ApJ...779..185K}, who define the observed virial parameter
of an individual structures as
\begin{equation}
\alpha_{obs} \equiv \frac{5\sigma_{tot}^{2}R}{GM}
\label{eq:ch3eq3}
\end{equation}
where $\sigma_{tot}$, $R$, and $M$ are the one-dimensional velocity
dispersion (including thermal and non-thermal gas motions), radius,
and mass of the structure, respectively. \citet{2013ApJ...779..185K}
summarize that the critical value of the virial parameter for a
spherical structure not supported by magnetic pressure is
$\alpha$~$\approx$~2. Structures with $\alpha$~$\ll$~2 are
supercritical and unstable against collapse unless they are supported
by magnetic fields. Structures with $\alpha$~$\gg$~2 are subcritical
and stable against collapse---they should dissipate unless they are
confined by high external pressure.  As discussed in
\citet{2013ApJ...779..185K}, this critical value includes the effects
of density gradients within and outside of the structure, as well as
surface pressure from the immediate surrounding medium.

For the virial parameter definition in Equation~\ref{eq:ch3eq3}, $M$
is the mass of each source derived from our \textit{Herschel} analysis
and is reported in Column~7 of Table~\ref{tbl:dendrodust}. A major
source of systematic uncertainty in the calculated mass is the
underestimate in the column density value derived from the
\textit{Herschel} SED fitting in Section~\ref{sec:ch3sec7sub1}. In
Section~\ref{sec:ch3sec7sub1}, we discussed how using a single
temperature model (ignoring the warmer foreground and background
component surrounding the colder L1451 region) for fitting the
\textit{Herschel} SEDs leads to estimated column densities that are
about a factor of two lower than the true column densities.  If the
measured column densities are a factor of two lower than the true
column densities, this directly leads to a mass underestimate by a
factor of two and a virial parameter overestimate by a factor of
two. Work has been done on the galactic scale to estimate dust mass
underestimation due to poor spatial resolution and temperature mixing
in \textit{Herschel} data. For example, \citet{2011A&A...536A..88G}
found that \textit{Herschel}-derived dust masses can be $\sim$50\%
underestimated in the Large Magellanic Cloud (LMC). Even though our
observations have much higher spatial resolution than observations of
the LMC, this is a useful comparison since the warm-cold temperature
contrast in local clouds like L1451 may be more extreme than on
galactic scales. For core scales, \citet{2009ApJ...696.2234S}
demonstrated that temperature variations along the line-of-sight will
lead to inaccurate measurements of dust properties from SED
fitting. We will consider the likely systematic underestimation of
mass from our SED fitting in the discussion of results below.

The radius of each structure, $R$, in the virial parameter definition is
half of the size reported in Column~4 of Table~\ref{tbl:dendrodust}.
The one-dimensional velocity dispersion of each structure,
$\sigma_{tot}$, in the virial parameter definition includes thermal
and non-thermal support against collapse. The thermal component of the
velocity dispersion was calculated as:
\begin{equation}
\sigma_{th} = \sqrt\frac{kT}{\mu m_{H}},
\label{eq:ch3eq4}
\end{equation}
where $T$ is the mean temperature of the dust within the structure,
$\mu$ is the mean molecular weight (2.33), and $m_{H}$ is the hydrogen
atomic mass. The
non-thermal component of the velocity dispersion was calculated for
each molecular tracer separately, as:
\begin{equation}
\sigma_{nth,tracer} = \sqrt{\sigma_{obs,tracer}^{2} - \sigma_{th,tracer}^{2}}
\label{eq:ch3eq5}
\end{equation}
where $\sigma_{obs,tracer}^{2}$ is the observed velocity dispersion of
the molecular emission within the structure, and $\sigma_{th,tracer}$
is the thermal velocity dispersion of the molecule at the mean
temperature found within the structure:
\begin{equation}
\sigma_{th,tracer} = \sqrt\frac{kT}{\mu_{tracer} m_{H}}.
\label{eq:ch3eq6}
\end{equation}
We determined $\sigma_{obs,tracer}^{2}$ by masking the CLASSy
molecular data cubes using the boundaries of each dust structure,
generating an integrated spectrum for each molecule, and fitting for
the velocity dispersion of that integrated spectrum.  The final
$\sigma_{total,tracer}$ value was then calculated as:
\begin{equation}
\sigma_{tot,tracer} = \sqrt{\sigma_{th}^{2} + \sigma_{nth,tracer}^{2}}
\label{eq:ch3eq7}
\end{equation}
and is reported in Columns~8, 9, and 10 of Table~\ref{tbl:dendrodust}
for \NtwoH{}, HCN, and \HCO{}, respectively. The temperature
overestimate from the single-temperature \textit{Herschel} SED fitting
discussed in Section~\ref{sec:ch3sec7sub1} systematically increases
the thermal component of the internal velocity dispersion of each
structure, thereby increasing the dispersive term used to calculate
$\alpha$. The temperature error is less significant than the
factor of two column density error discussed above, but still
contributes a $\sim$7\% systematic increase in the virial parameter if
the measured temperature is overestimated by 2.5~K from the true
value.

Virial parameters for each dust leaf structure are reported in
Table~\ref{tbl:virialdust}, with unique values for each molecular
tracer resulting from different velocity dispersions between the
tracers. For each tracer, we first report the measured virial
parameter using the masses derived from the \textit{Herschel} SED
fitting, and then report a value reduced by a factor of two to account
for the likely systematic underestimation of the fitted column density
values discussed above.

Two things are clear from the $\alpha$ values in
Table~\ref{tbl:virialdust}. First, the $\alpha$ values typically vary
by a factor of two between tracers, with \NtwoH{} and \HCO{} data
giving the lowest and highest virial parameter, respectively. This
trend with molecular tracer is not surprising, since it connects to
the differences in spatial extent and kinematics between molecular
emission discussed in Sections~\ref{sec:ch3sec4} and
\ref{sec:ch3sec5}.  The \HCO{} traces the most spatially extended
emission on the plane of the sky, implying it is tracing the most
extended emission along the line-of-sight if the two spatial
directions are correlated. This implication is strongly supported by
\HCO{} having the largest line-of-sight velocity dispersions. Since
the virial parameter scales as the square of the velocity dispersion
of each molecule, the larger \HCO{} velocity dispersion leads to the
larger virial parameter. This shows that the choice of kinematic
tracer has large impact on the interpretation of a given structure.

Second, all virial parameter values are greater than two if we use the
masses derived from the single-temperature model of \textit{Herschel}
SED fitting. If we assume that the mass is underestimated by a factor
of two and the virial parameter is overestimated by a factor of two
(ignoring the systematic temperature uncertainty also increasing the
virial parameter), then some of the structures have virial parameters
near or less than the critical value of 2 for instability against
collapse. Structures 1, 3, 4, and 7 have $\alpha$ above two regardless
of the molecular tracer used or assumption of mass. These structures
have the weakest dense gas emission in the field and/or appear the
least centrally condensed in the column density maps, so it is
reasonable that they appear to be the least unstable against collapse.
Structures 0, 2, 5, and 6 have $\alpha$ near or below two, appear
centrally condensed in the column density map, and have strong
molecular emission in the line maps. Structure 0 contains Per~Bolo~4,
structure 2 contains Per~Bolo~6, structure 5 contains L1451-west, and
structure 6 contains L1451-mm. These results together support these
structures (or at least the central parts of these structures) being
unstable against collapse, and the L1451 region as a whole being at
the onset of forming a few protostars.

We will extend this type of virial analysis in upcoming work to
compare virial parameters across CLASSy regions. This way, even if
there are systematic and statistical uncertainties, all structures
will have been observed and analyzed in the same way, making a
relative comparison of virial parameters across cloud regions at
different stages of evolution possible.

\begin{deluxetable}{c c c c c c c}
\tabletypesize{\small}
\tablecaption{Virial Parameters}
\tablewidth{0pt}
\tablehead
    {
      \hline
      & \multicolumn{2}{c}{$\alpha$ (\NtwoH{})} & \multicolumn{2}{c}{$\alpha$ (HCN)}  & \multicolumn{2}{c}{$\alpha$ (\HCO{})} \\
      \cline{2-7}
       \colhead{{No.}} & \colhead{{1-temp}} & \colhead{{2-temp$^{a}$}} & \colhead{{1-temp}} & \colhead{{2-temp$^{a}$}} & \colhead{{1-temp}} & \colhead{{2-temp$^{a}$}} 
}
    \startdata
    (1) & (2) & (3) & (4)   & (5) & (6) & (7)\\
    \hline
    0   &  3.0   & 1.5 &  4.0 & 2.0 & 6.1 & 3.1 \\
    1   &  ...   & ... &  28.5 & 14.3 & 32.9  & 16.5 \\
    2   &  2.3   & 1.2 &  2.8  & 1.4 & 5.2 & 2.6 \\
    3   &  15.3  & 7.7 & 13.1 & 6.6  & 26.0 & 13.0 \\
    4   &  ...   & ... &  17.5 & 8.8 & 24.8 & 12.4 \\
    5   & 3.8    & 1.9 &  9.3 & 4.7  & 13.5 & 6.8 \\
    6$^{b}$&  4.4 & 2.2  &  4.5 & 2.3 & 7.9 & 4.0 \\
    7   & 6.2    &  3.1 & 6.8  & 3.4 & 14.0 & 7.0 \\
\enddata
\vspace{-0.5cm} \footnotesize \tablecomments{(1) Leaf number from
  Table~\ref{tbl:dendrodust}. (2) Virial parameter calculated using
  Equation~\ref{eq:ch3eq3} with the \NtwoH{} velocity dispersion and
  mass derived from \textit{Herschel} single-temperature SED
  fitting. (3) Column 2, corrected by a factor of two to account for
  the likely underestimation of column density (and therefore mass)
  from single-temperature SED fitting discussed in text. (4) Same as
  Column 2, but using the HCN velocity dispersion. (5) Same as Column
  3, but using the HCN velocity dispersion. (6) Same as Column 2, but
  using the \HCO{} velocity dispersion. (7) Same as Column 3, but
  using the \HCO{} velocity dispersion.  $^{a}$We are not formally
  using a two-temperature model to fit the \textit{Herschel} SEDs, but
  are applying a correction factor that estimates the likely effect of
  using such a model. $^{b}$L1451-mm is located within this
  structure.}
\label{tbl:virialdust}
\end{deluxetable}

\clearpage

\subsection{Closer Look at L1451-mm and L1451-west}
\label{sec:ch3sec8sub3}
In this section, we explore the properties of the two centrally
condensed, roughly spherical cores in L1451. One is L1451-mm, which we
know is a compact object containing a YSO or a FHSC
\citep{2011ApJ...743..201P}. The other is L1451-west, which has been
discovered with these observations. We summarize the morphology and
kinematics of L1451-mm below, and then describe the properties of
L1451-west and how they compare to L1451-mm.

\begin{figure}[!h]
\centering \includegraphics[scale=1.0]{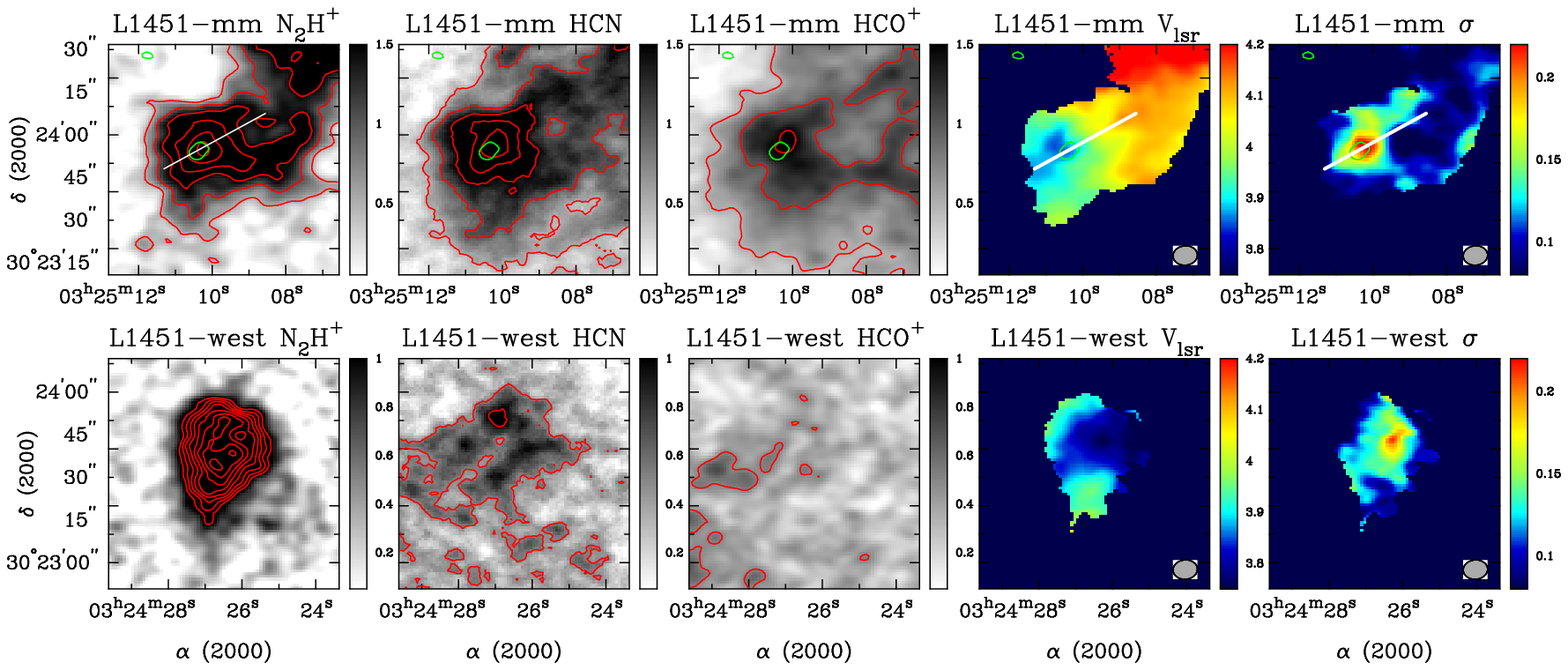}
\caption{\small The top and bottom rows show properties of L1451-mm
  and L1451-west, respectively. From left-to-right, the panels are:
  \NtwoH{}, HCN, and \HCO{} integrated intensity (\Jybmkms{}),
  \NtwoH{} centroid velocity (\kms{}), and \NtwoH{} velocity
  dispersion (\kms{}). The kinematic maps of each source are on the
  same color scale.  The velocity dispersion is expressed as Gaussian
  $\sigma$ (FWHM/2.355 in \kms{}).  The green 0.0039 m\Jybm{} contour
  represents our L1451-mm compact continuum detection. The red
  contours are used to accentuate specific features in the greyscale
  maps. The white lines represent the direction of the measured
  velocity gradient through L1451-mm.}
\label{fig:conts}
\end{figure}

The top row of Figure~\ref{fig:conts} shows molecular and continuum
features of L1451-mm. L1451-mm is the only confirmed compact continuum
core in L1451, and all molecules trace strong emission near, and
surrounding its location.  All molecules peak in integrated emission
at a location slightly offset from the L1451-mm compact continuum
core. The molecular emission that surrounds the core is more
concentrated for \NtwoH{}, and more widespread for HCN and \HCO{}.
The \NtwoH{} centroid velocity field shows a gradient across the core,
which \citet{2011ApJ...743..201P} modeled as a rotating and infalling
envelope using slightly higher angular resolution (5\arcsec{}) data;
we measure a gradient of about 6~\kmsppc{} through the peak of
integrated emission at a position angle of 120~deg east of north (see
white line in the top row panels of Figure~\ref{fig:conts}). The
velocity dispersion field is narrowest around the edges of the core
with minima $\sim$0.07~\kms, and it shows an increase toward the core
location, with a peak $\sim$0.2~\kms.  See \citet{2011ApJ...743..201P}
for detailed discussion and modeling of the possibility that L1451-mm
is a dense core with a central YSO and disk, or a dense core with a
central FHSC.

The bottom row of Figure~\ref{fig:conts} shows molecular and continuum
features of L1451-west.  The \NtwoH{} integrated intensity map does
not have a single peak of integrated emission at the center of the
structure. Instead, there are two peaks in the southern half of the
source $\sim$2.5~\Jybmkms, and one peak in the northern half of the
source $\sim$2.4~\Jybmkms{} (the red contours in the bottom-left panel
of Figure~\ref{fig:conts} represent 1.1--2.5~\Jybmkms{} in
0.2~\Jybmkms{} increments). The source is considerably weaker in HCN,
and not detected in \HCO{}. This apparent lower-abundance of \HCO{}
may indicate that this might be a very cold, dense region of L1451,
where CO is depleted. If CO is frozen out onto dust grains, then it is
not able to destroy \NtwoH{}, and \NtwoH{} remains abundant. Likewise,
if CO is frozen out, it is not available to create \HCO{} through
collisions with H$_{3}$$^{+}$ \citep{1980ApJS...43....1P}. Our
temperature map in Figure~\ref{fig:herschelT} does not suggest that
dust around L1451-west is significantly colder than L1451-mm, but
higher-resolution continuum observations with modeling to remove any
warm component could reveal differences in the two sources that are
not resolvable with the current data. It is also possible that the
\HCO{} ($J=1\rightarrow0$) photons are being absorbed by \HCO{} molecules
in lower-density foreground gas; observations of higher-J transitions
are needed to test this.

The \NtwoH{} centroid velocity field of L1451-west shows a relatively
complex structure compared with L1451-mm, with higher velocity
emission in the southern and northeastern sections of the core, and
lower velocity emission toward the center and west of the core.  The
velocity dispersion field of L1451-west is narrowest around the edges
of the core with minima $\sim$0.07~\kms, and it peaks near
$\sim$0.2~\kms{} at the location of lowest velocity emission in the
central part of the source. It is possible that infall is producing
the increased velocity dispersion, in addition to any undetected
outflows and/or unresolved disk rotation from source center. Infall
would broaden the molecular emission toward the core center, with the
blueshifted emission being brighter than the redshifted emission if
the \NtwoH{} emission is optically thick. This could explain why the
peak of velocity dispersion is at the location with the bluest
centroid velocity. The signal-to-noise of the detectable HCN spectra
are too low to look for evidence of infall motions, so this needs to
be followed up with deeper observations of optically thick and thin
lines.

There is no compact continuum emission detected toward L1451-west. The
3$\sigma$ flux density limit for a point source ($\lesssim$3\arcsec{}
= 700~AU) in our observations is 3.9~mJy, corresponding to a mass
limit of 0.08~\Msun{} for the conversion from mJy to \Msun{} discussed
in Paper~I. We use the location of peak velocity dispersion as a proxy
for the location of a compact continuum core, if it exists; this peak
is offset from the peak in the \textit{Herschel}-derived column
density map by about 16\arcsec{}.

We estimated the size and physical density of L1451-mm and L1451-west
for a more detailed comparison. To determine the size, we used the
MIRIAD {\tt imfit} routine to fit a two-dimensional Gaussian to the
\NtwoH{} integrated intensity map.  The geometric means of the major
and minor axes are 37\arcsec{} and 32\arcsec{} for L1451-mm and
L1451-west, respectively. To determine the maximum physical density,
we took the peak column density of each source within the fitted
Gaussian, and assumed their depth was the same extent as their
geometric mean across the plane of the sky.  For both sources, the
physical density is 7--8$\times$~10$^{4}$~cm$^{-3}$ in the 8\arcsec{}
beam. This density is sensible, considering that we observe \NtwoH{}
and HCN ($J=1\rightarrow0$) in L1451, which are molecular transitions
with effective excitation densities on the order of
10$^{5}$~cm$^{-3}$.

We also compared the mass within 4200~AU (18\arcsec{} at $d$=235~pc)
to the intrinsic radius at 70\% of peak intensity, following the
analysis of starless cores presented in
\citet{2008A&A...487..993K}. We measured the mass using the
\textit{Herschel} data, and got 0.22 and 0.23~\Msun{} for L1451-mm and
L1451-west, respectively. We measured the radius using the \NtwoH{}
integrated intensity, since the dust data is not as well resolved to
derive accurate radius measurements. L1451-mm measured 8\arcsec{}, and
L1451-west measured 12\arcsec{} at 70\% of the peak
intensity. Compared to the sample of starless cores in
\citet{2008A&A...487..993K}, L1451-mm is more centrally dense than
most starless objects, while L1451-west is near the upper limit of
central density seen in starless objects.

All of these observational results point to L1451-west being similar
to, but slightly less evolved than, L1451-mm, yet more evolved than a
prestellar core.  Although deep continuum and spectral line
observations will be needed to determine the true nature of these
sources, they are some of the best evidence that L1451 is a region
that is just starting to form its first stars.

\subsection{Three-Dimensional Morphology of L1451}
\label{sec:ch3sec8sub4}
Comparing the projected size of cloud structure to its depth along the
line-of-sight gives a better understanding of the three-dimensional
geometry of a region where stars are forming (e.g., a region that is
primarily planar/sheet-like versus one that is more spherical). We
described a statistical method to estimate the typical line-of-sight
depth of cloud structures in Paper~I. The method uses the spatial and
kinematic properties of dendrogram objects presented in
Tables~\ref{tbl:dendrohcop}--\ref{tbl:dendron2hp}.  It assumes that
the \vlsr{} variation ($\Delta$\vlsr; Column 11) of a structure scales
with its projected size (Column 9) in a turbulent medium.  It also
assumes that the mean non-thermal velocity dispersion ($\langle \sigma
\rangle_{\mathrm{nt}}$) of a dendrogram structure scales with its
depth along the line-of-sight in a turbulent medium. We calculate
$\langle \sigma \rangle_{\mathrm{nt}}$ for each structure in all three
molecular tracers by subtracting the thermal velocity dispersion of
10~K gas of the given tracer from the value reported in Column~12 of
Tables~\ref{tbl:dendrohcop}--\ref{tbl:dendron2hp}: $\langle \sigma
\rangle_{\mathrm{nt}} = \sqrt[]{\langle \sigma \rangle^{2} -
  \sigma_{\mathrm{th}}^{2}}$. With those assumptions, we create two
size-linewidth relations using all the dendrogram structures (one with
projected size versus $\langle \sigma \rangle_{\mathrm{nt}}$, and the
other with projected size versus $\Delta$\vlsr), and take the spatial
scale where they cross as the typical depth of the region. See Paper~I
for the theoretical framework and numerical results that justify this
method.

\begin{figure}[!h]
\centering \includegraphics[scale=0.7]{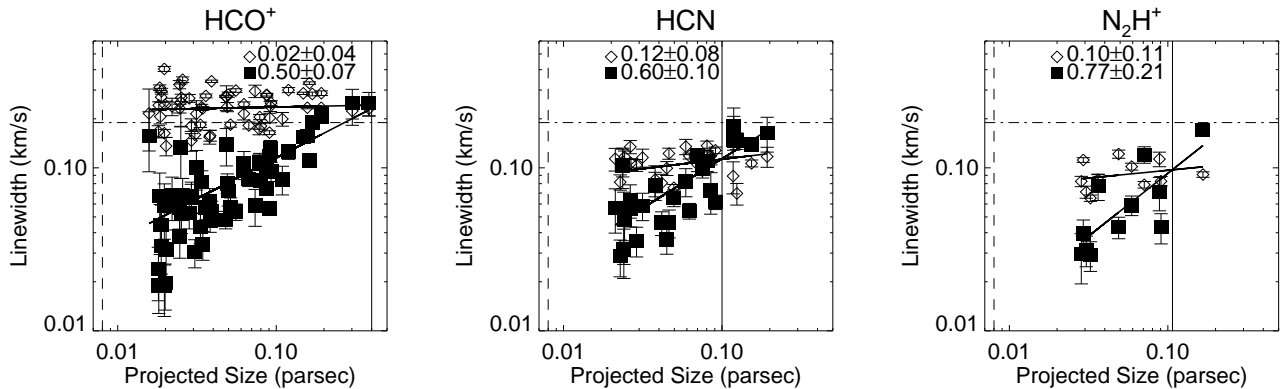}
\caption{ \small Scaling relations between projected structure size
  and \vlsr{} variation ($\Delta$\vlsr; solid squares), and projected
  structure size and mean non-thermal velocity dispersion ($\langle
  \sigma \rangle_{\mathrm{nt}}$; open diamonds), for each
  molecule. The solid lines represent single power-law fits to the
  data points. The horizontal line represents the typical thermal
  speed for H$_{2}$ at gas kinetic temperatures near 10~K. The
  vertical dashed line represents our spatial resolution of
  $\sim$0.008~pc. The vertical solid line represents the spatial scale
  where the power-law fits intersect.}
\label{fig:szlw}
\end{figure}

We used the method in previous papers to argue that the typical depth
of the \NtwoH{} emission in the CLASSy Barnard~1 and Serpens Main
regions was 0.1--0.2~pc (see Paper~I and \citet{2014ApJ...797...76L}).
Figure~\ref{fig:szlw} shows the size-linewidth relations for each
molecule. As in the analysis of Barnard~1 in Paper~I and Serpens Main
in \citet{2014ApJ...797...76L}, the projected size-$\Delta$\vlsr{} for
each molecule has a positive slope (which is expected for a turbulent
medium), while the projected size-$\langle \sigma
\rangle_{\mathrm{nt}}$ relationship has a flatter slope (which is
expected if all structures have a similar depth along the
line-of-sight, independent of their projected size). The best-fit
lines to the relations cross at a size-scale of 0.11 and 0.10~pc for
\NtwoH{} and HCN, respectively, indicating that those molecules are
tracing structures $\sim$0.1~pc in depth. The \HCO{} fits cross near
0.40~pc, indicating that it is tracing structures with larger
line-of-sight depths compared to the other molecules. This is
consistent with \HCO{} detecting larger-scale, lower density gas than
the other molecules.

These results show that the \NtwoH{} emission in L1451 has similar
depth as it does in Barnard~1 and Serpens Main. However, there are
differences between L1451 and the other regions that lead us to
different conclusions about the large-scale structure of L1451.  In
Paper~I, we said that the depth of Barnard~1 (0.1--0.2~pc) is
comparable to the largest size of individual \NtwoH{} dendrogram
structures identified from the isolated \NtwoH{} hyperfine component
(0.2--0.3~pc). But we followed up by discussing how the Barnard~1 has
contiguous \NtwoH{} structure at parsec-scales when considering the
full \NtwoH{} emission instead of just the isolated hyperfine
component. We used the estimated 0.1--0.2~pc depth and observed parsec
projected size of Barnard~1 to conclude that the region is flattened
at the largest detectable scales.

In projection, the full \NtwoH{} emission, sub-millimeter continuum
emission, and other molecular emission from L1451 region does not
appear to have contiguous structure across parsec scales like all
other CLASSy regions. Instead, the emission is concentrated in a few
major features, sub-parsec in size, that were identified in
Section~\ref{sec:ch3sec4}. Because of this, we argue that the L1451 is
not a flattened, sheet-like region of dense gas and dust at
parsec-scales with connected substructure. In \NtwoH{}, it appears
more like a loose collection of dense concentrations that are
$\sim$0.2~pc projected on the sky and $\sim$0.1~pc deep.

The physical density of the regions of L1451 with \NtwoH{} emission
can be estimated from the cloud depth and column density.  For a
0.1~pc depth of \NtwoH{} emission and a mean N(H$_{2}$) of
6~$\times$~10$^{21}$~cm$^{-2}$ measured from the \textit{Herschel}
data, the derived physical density is 2~$\times$~10$^{4}$~cm$^{-3}$ in
the regions of L1451 with \NtwoH{} emission. It is possible to excite
\NtwoH{} at a range of physical densities, where the range depends on
the molecular column density (and thus relative abundance of \NtwoH{}
to H$_{2}$), the gas kinetic temperature, and the linewidth of
emission. Using the RADEX program \citep{2007A&A...468..627V}, we find
that \NtwoH{} ($J=1\rightarrow0$) can reach 1~K brightness
temperatures with a physical density of
2.7~$\times$~10$^{4}$~cm$^{-3}$ if the gas kinetic temperature is
12~K, the molecular column density of \NtwoH{} is
6~$\times$~10$^{12}$~cm$^{-2}$, and the linewidth of emission is
0.3~\kms. This physical density is essentially the same as derived
from our depth and column density measurements, providing a
consistency check to our measurements.

We can compare the typical depth of the molecular emission to the
projected size of the largest-scale structures of each molecule in
Figure~\ref{fig:dendrobranchmap} to infer the three-dimensional
morphology of the individual molecular structures.  The projected size
of \NtwoH{} structure 21 (lowest level branch of Features~A) is
$\sim$0.17~pc, structure 18 (lowest level branch of Features~C) is
$\sim$0.09~pc, and structure 17 (lowest level branch of Features~B) is
$\sim$0.17~pc.  For a typical \NtwoH{} depth of 0.11~pc, these
structures have axis ratios of 1.5:1 and 0.8:1, with a mean of 1.3:1.
The projected size of HCN structure 45 (lowest level branch of
Feature~A) is $\sim$0.24~pc, structure 43 (lowest level branch of
Feature~C) is $\sim$0.17~pc, and structure 40 (lowest level branch of
Feature~B) is $\sim$0.19~pc.  For a typical HCN depth of 0.10~pc,
these structures have axis ratios between 2.4:1 and 1.7:1, with a mean
axis ratio of 2.0:1.  The projected size of \HCO{} structure 111
(lowest level branch connecting Features A and C) is $\sim$0.43~pc,
and structure 112 (lowest level branch connecting Features B, H, F, G,
E, and I) is $\sim$0.46~pc. For our derived typical \HCO{} depth of
0.40~pc, these structures have axis ratios of 1.1:1 and 1.2:1.

With axis ratios between 0.8:1 and 2.4:1 for individual structures
that do not connect at larger scales, these results support the L1451
region being composed of discrete, high-density structures that are
approximated as ellipsoids.  In addition to not having contiguous,
flattened structure at parsec scales like the other CLASSy regions
studied to-date (Serpens Main and Barnard~1), none of the large-scale
structures of L1451 appear to have clear filamentary substructure like
was the case in all other CLASSy regions. These results differentiate
L1451 from the other regions in a way beyond the lack of protostars
and outflows.

If L1451 is not a flattened, sheet-like region at parsec-scales, if it
does not have prominent filamentary structure, and if it has less
active star formation than the other CLASSy regions, then it is
natural to speculate what caused these differences. A cloud complex,
like Perseus, that spans tens of parsecs will have different turbulent
energies in different parts of the cloud. Simulations of
turbulence-driven star formation with supersonic turbulence
\citep[e.g.,][]{2012ApJ...761..156F} show that star formation within
several-parsec scale clouds is clustered in regions where material has
been compressed to high densities, leaving voids of star formation in
other parts of the cloud. We speculate that the L1451 region of
Perseus may not have been compressed as strongly by supersonic
turbulence to form an overdense sheet-like structure at parsec-scales
like may have happened several parsecs away near the cloud regions
that became Barnard~1 and NGC~1333. Without a comparable push to
high-density that other more active regions of Perseus got, the L1451
region may have been predisposed to forming fewer stars.

\section{Summary of L1451 Results}
\label{sec:ch3sec9} 

We presented observations and analysis of the L1451 region of the
CARMA Large Area Star Formation Survey. We summarize the key findings
below. 

\begin{enumerate}

\item Only one compact continuum source is detected at 3~mm, down to a
  3$\sigma$ limit of 3.9~m\Jybm{} (0.08~\Msun{} limit) in a
  9.2\arcsec{}~$\times$~6.6\arcsec{} beam. The detected source is
  L1451-mm, which has previously been identified as a FHSC candidate.

\item We detect widespread \HCO{}, HCN, and \NtwoH{}
  ($J=1\rightarrow0$) emission in L1451. The \HCO{} emission covers
  the largest area of the cloud, which we attribute to \HCO{} tracing
  lower-density material than the other molecules. HCN emission
  morphology is nearly identical to \HCO{}, although it is weaker and
  less spatially extended in most regions. We detect no molecular
  outflows from \HCO{} or HCN. The \NtwoH{} emission appears the most
  spatially compact, and reveals a newly identified compact core,
  which we call L1451-west.
    
\item All molecules trace gas at similar systemic velocities, but the
  velocity dispersion of the \HCO{} emission is significantly larger
  than that of HCN or \NtwoH{}: mean dispersions are 0.29, 0.16, and
  0.12~\kms, respectively. This is likely due to \HCO{} tracing
  lower-density gas at larger scales than the other two molecules; in
  a turbulent medium, there is more power on larger scales, which will
  produce larger velocity dispersions.

\item We derive column density and temperature maps from
  \textit{Herschel} observations at 160, 250, 350, and 500~$\mu$m. The
  values were derived using modified blackbody spectrum fits to the
  data, and column densities agree to within 5\% of \textit{Planck}
  results when compared at the same angular resolution. The
  temperatures toward the densest regions are $\sim$2--3~K warmer than
  kinetic temperatures derived from single-pointing ammonia
  observations toward dense cores. We attribute this difference to a
  limitation of using a single-component fit when modeling SEDs from
  cold, dense regions of molecular clouds. A simple multi-layer model
  shows that having warm, 17~K foreground and background emission
  surrounding cold, 9~K cloud emission, can cause temperatures to be
  overestimated by a few~K, and column densities to be underestimated
  by a factor of two.
  
\item The structure of the star-forming material in L1451 that is
  traced by the cumulative molecular emission we detect with CLASSy is
  very similar to the column density structure we derived from
  \textit{Herschel} observations. All of the cloud locations that are
  above the \Av{}~$\sim$~8 threshold for star formation are detected
  in dense gas, and 90\% of the molecular emission is at column
  densities above 1.9~$\times$~10$^{21}$~cm$^{-2}$. This shows that
  high-resolution observations of this suite of spectral lines over
  large areas of molecular clouds are an excellent probe of the
  structure and kinematics of star forming material in young regions.

\item We use our non-binary dendrogram algorithm to identify dense gas
  structures in the \HCO{}, HCN, and \NtwoH{} data cubes.  The number
  of identified leaf and branch dendrogram structures decreases from
  \HCO{} to HCN to \NtwoH{}, providing an apparently wide range of
  hierarchical complexity in the different tracers. However, a uniform
  comparison of tree statistics that accounts for differences in the
  noise-level of each data cube shows that all tracers are identifying
  structures that are fragmenting in a similar way. The differences in
  dendrogram branching levels and number of structures between tracers
  are likely due to differences in density sensitivity of the
  different tracers.  We show that tree statistics of the gas
  structure surrounding L1451-mm (a confirmed young protostar or FHSC)
  is very similar to that of the gas surrounding Per~Bolo~6 (a
  single-dish continuum detection), and argue this is an indication
  that star formation is proceeding in a similar fashion in both
  regions, with Per~Bolo~6 a likely site of future star formation.

\item A virial analysis of dust structures identified in our
  \textit{Herschel}-derived extinction map of L1451 resulted in all
  structures being stable against collapse when using the masses
  derived from single-temperature \textit{Herschel} SED fitting and
  velocity dispersion from any CLASSy molecular tracer.  However, we
  discussed how the virial parameter is likely overestimated by about
  a factor of two from the systematic underestimation of column
  densities from the simplistic single-temperature SED fitting. Also,
  the virial parameter of a given structure is strongly dependent on
  the molecular tracer used to estimate its kinetic energy; \NtwoH{}
  and \HCO{} resulted in the lowest and highest virial parameters,
  respectively, with typical differences of about a factor of two.
  The dust structures that appear most centrally condensed in the dust
  maps are near or below the critical value for instability when
  considering the systematic uncertainty to the mass and using the
  \NtwoH{} or HCN velocity dispersion.

\item We detect two strong, centrally condensed \NtwoH{} structures:
  L1451-mm, and a newly identified source that we label
  L1451-west. L1451-mm was characterized by
  \citet{2011ApJ...743..201P} as a FHSC candidate or young
  protostar. Our data shows that L1451-west is similar to, but likely
  younger than L1451-mm.  It has strong emission from \NtwoH{} but
  appears depleted in \HCO{}, unlike L1451-mm which has strong
  emission from both molecules. Both sources show \NtwoH{} velocity
  dispersions peaking at the core center.  L1451-west is less
  centrally condensed than L1451-mm, but more centrally condensed that
  the typical prestellar core; this indicates that L1451-west is
  likely at an evolutionary state between the typical prestellar core
  and a FHSC or young protostar. Follow-up observations will determine
  if L1451-west is a viable FHSC candidate.

\item We use our size-linewidth analysis first presented in Paper~I to
  estimate the typical line-of-sight depth of dense gas structures in
  L1451.  Typical inferred line-of-sight depths for \HCO{} structures
  are 0.40~pc, and for HCN and \NtwoH{} structures are
  $\sim$0.10~pc. These depths are found to be within a factor of two
  of the projected sizes of gas structures, showing that the
  three-dimensional dense gas morphology in L1451 is relatively
  ellipsoidal at the largest detectable scales when compared to the
  more evolved CLASSy regions that appear flattened/sheet-like at the
  largest scales.

\end{enumerate}
  
Overall, these observations support the creation of complex,
hierarchical dense gas structure in molecular clouds without internal
feedback from protostars.  We cannot say that individual structures
within L1451 are definitively gravitationally bound based on our
analysis, but that several structures are consistent with being
unstable to collapse within the uncertainties in the data; we also
know at least one star is forming in the region based on compact
continuum detections. The velocities in these structures are larger
than pure thermal and range from subsonic gas motions on the smallest
scales (as traced by \NtwoH{}) to slightly supersonic gas motions on
the largest scales (as traced by \HCO{}); this supports external
supersonic turbulence on even larger scales than those traced by
\HCO{} being the driver of the first structure within the densest
regions of molecular cloud complexes; those structures then become the
dense, fertile grounds for the future formation of stars.

The science-ready spectral line data cubes for L1451 are hosted at our
project website: \\ http://carma.astro.umd.edu/classy. We welcome the
community to make use of the data.

\acknowledgments The authors would like to thank the referee for
encouraging critiques that improved the paper, and all members of the
CARMA staff that made these observations possible. CLASSy was
supported by AST-1139990 (University of Maryland) and AST-1139950
(University of Illinois).  Support for CARMA construction was derived
from the Gordon and Betty Moore Foundation, the Kenneth T. and Eileen
L. Norris Foundation, the James S. McDonnell Foundation, the
Associates of the California Institute of Technology, the University
of Chicago, the states of Illinois, California, and Maryland, and the
National Science Foundation. Ongoing CARMA development and operations
are supported by the National Science Foundation under a cooperative
agreement, and by the CARMA partner universities.

 {\it Facility:} \facility{CARMA}

\bibliography{mybib_b1}

\appendix
\section{Noise Dependence of Dendrogram Structure and Statistics}
\label{app:appendixC}

In Paper~I, we analyzed a single non-binary dendrogram from the
position-position-velocity (PPV) cube of the isolated hyperfine
component of \NtwoH{} in Barnard~1. In this paper, we analyzed the
non-binary dendrograms from the PPV cubes of the strongest hyperfine
components of \NtwoH{} and HCN, and the single component of \HCO{} in
L1451. There was a noise difference of 15\% between those cubes, and
the goal of this section is to establish how we can compare dendrogram
structure and statistics of data with different noise-levels.  We will
argue that the key for a proper comparison is to construct the
dendrograms using absolute, physical units (as opposed to units based
on the sensitivity of an individual data set) for certain dendrogram
algorithm parameters, including the parameters for: 1) the minimum
height for a local maximum to peak above the level where it merges
with adjacent local maxima to be considered a real leaf ({\tt
  minheight}), and 2) the minimum allowed branching step height
between adjacent levels in the dendrogram ({\tt stepsize}).

We will demonstrate those points using two integrated intensity maps
of Barnard~1\footnote{We use Barnard~1 instead of L1451 for
  demonstration in this appendix because Barnard~1 was the first
  CLASSy region analyzed and used to test the dendrogram analysis
  methods.} that have similar signal but different noise, created from
different hyperfine components in the \NtwoH{} PPV cube presented in
Paper~I. The \NtwoH{} ($J=1\rightarrow0$) transition is split into
seven resolvable hyperfine components, which have effectively the same
velocity dispersion and excitation condition but different peak
intensity. We can integrate over different combinations of hyperfine
components to create moment maps with different signal-to-noise ratio
(SNR). Specifically, we integrate over all seven hyperfine components
to create a ``full-velocity component'' integrated intensity map
(referred to as the FVC map later in the section), and over the
isolated velocity component to make the ``lowest-velocity component''
integrated intensity map (referred to as the LVC map later in the
section). The LVC map has lower signal than the FVC map, but also
lower noise ($\sigma_{LVC}$) in emission-free regions compared to the
FVC map ($\sigma_{FVC}$), since noise scales as $\sqrt{N}\sigma$,
where N is the number of channels, and $\sigma$ is the noise in a
single channel.

The questions we want to answer hinge on how to consistently analyze
dendrogram structure and statistics when we change the noise in a map
while keeping the signal the same. Therefore, we normalize the FVC map
to have the same signal as the LVC map in regions where the LVC map
has emission greater than 6$\sigma_{LVC}$. In regions less than
6$\sigma_{LVC}$ in the LVC map, we scale the FVC map down by a factor
of 7, which is the mean normalization factor in regions of strong
signal. We then calculate the rms of the normalized FVC map (referred
to as the NFVC map later in the section), $\sigma_{NFVC}$, which is
0.03~\Jybmkms{}, compared to $\sigma_{LVC}$ of 0.09~\Jybmkms{}. The
result is a NFVC map with similar signal to the LVC map with three
times lower noise; this is the scenario we want to test our dendrogram
algorithm on.

We ran our non-binary dendrogram algorithm on the Barnard~1 versions
of these two maps in two different ways. First, we allowed each
dendrogram to have a {\tt stepsize} and {\tt minheight} based on the
rms of each map, meaning {\tt stepsize}~=~1$\sigma_{NFVC}$ and {\tt
  minheight}~=~2$\sigma_{NFVC}$ for the NFVC map, and {\tt
  stepsize}~=~1$\sigma_{LVC}$ and {\tt minheight}~=~2$\sigma_{LVC}$
for the LVC map. Second, we forced each dendrogram to have the same
{\tt stepsize} and {\tt minheight} based on the noisier map, meaning
{\tt stepsize}~=~1$\sigma_{LVC}$ and {\tt
  minheight}~=~2$\sigma_{LVC}$ for both maps. Using absolute units
across maps as in this second case could be beneficial because
absolute units are linked with physical properties of a cloud, while
noise units are linked with the properties of the individual
observations. For example, if a study wants to link fragmentation
properties of column density structure in different clouds to the
properties of star formation in those clouds, it would be more
sensible to define an absolute unit, such as a minimum mass or column
density that is always the same across clouds, for branching and leaf
identification, rather than a noise unit, such as 1$\sigma$, which
can be different for different clouds.

\begin{figure}[!h]
\centering \includegraphics[scale=0.095]{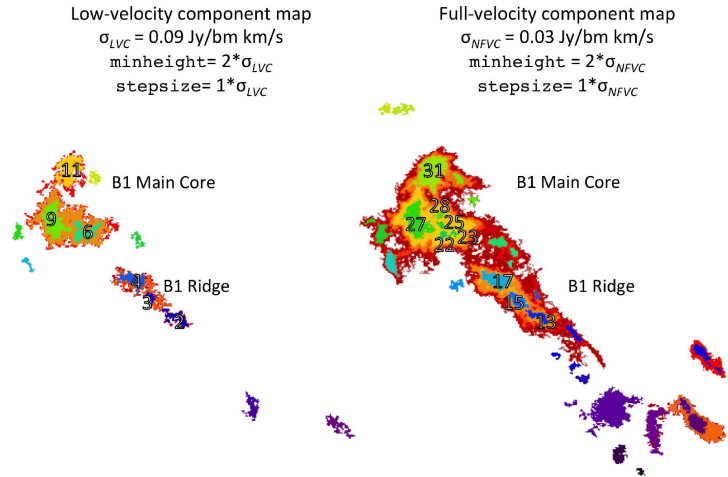}
\caption{\small Structures identified by our non-binary dendrogram
  algorithm from the Barnard~1 integrated intensity maps made from the
  lowest-velocity hyperfine component (left) and from all hyperfine
  components normalized as discussed in the text (right). The
  dendrogram algorithm parameters are listed in the figure, and are
  based on the noise level of the individual maps. Structures are
  labeled with numbers that correspond to leaves and branches in
  Figure~\ref{fig:b1dendrocomparetree1}. See Paper~I for a description
  and analysis of the Barnard~1 region.}
\label{fig:b1dendrocompare1}
\end{figure}

\begin{figure}[!h]
\centering 
\includegraphics[scale=0.095]{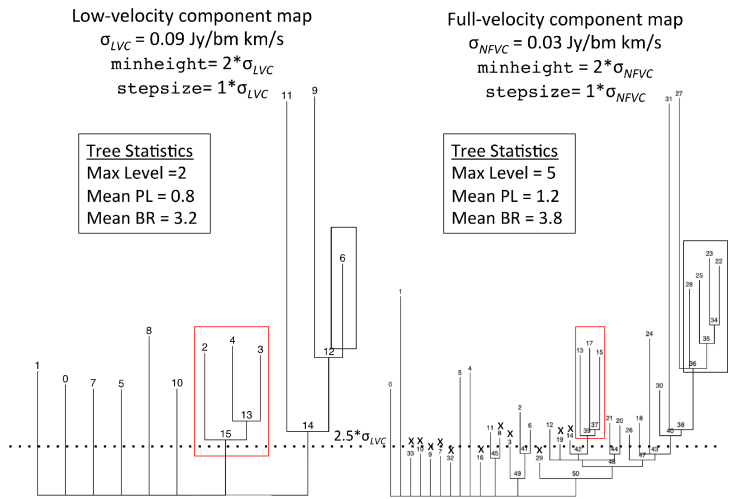}
\caption{\small Non-binary dendrograms from the integrated intensity
  maps made from the lowest-velocity hyperfine component (left) and
  from all hyperfine components normalized as discussed in the text
  (right) using the dendrogram algorithm parameters listed in the
  figure. The boxes highlight examples of the same regions of emission
  between the two maps for a comparison of identified structures when
  using non-uniform algorithm parameters on maps with different noise
  levels. The tree statistics in the figure inset are discussed in the
  text.  See Paper~I for a description and analysis of the Barnard~1
  region.}
\label{fig:b1dendrocomparetree1}
\end{figure}

For the first case, where we allowed the {\tt stepsize} and {\tt
  minheight} parameters to be based on the noise-level of the
individual map, the resulting dendrogram-identified structures for the
two Barnard~1 maps are shown in Figure~\ref{fig:b1dendrocompare1}, and
the dendrograms are shown in Figure~\ref{fig:b1dendrocomparetree1}.
Two of the strongest peaks in the region are similarly identified in
both maps (structures 9 and 11 in the LVC map are 27 and 31 in the
NFVC map).  However, some leaves break up into multiple structures
when lowering the noise of the map and also lowering the absolute
threshold for creating a leaf and a branching step (structure 6 in the
LVC map becomes structures 22, 23, 25, and 28 in the NFVC map).

This shows that the hierarchical structure of two maps of the same
region that have different noise can significantly change if {\tt
  minheight} and {\tt stepsize} are kept in terms of the sensitivity
of each individual map. This makes sense if we consider an extreme
limit of very low noise maps that may soon be produced by ALMA. If the
noise in a map is extremely reduced, structures peaking 2$\sigma$
above a merger level with other structures would likely not represent
physically relevant features, but small-scale variations on top of
physical meaningful features. In this case of extremely low noise, it
would make more sense to define a physically relevant minimum unit for
studying structure, such as a minimum mass.

\begin{figure}[!h]
\centering \includegraphics[scale=0.095]{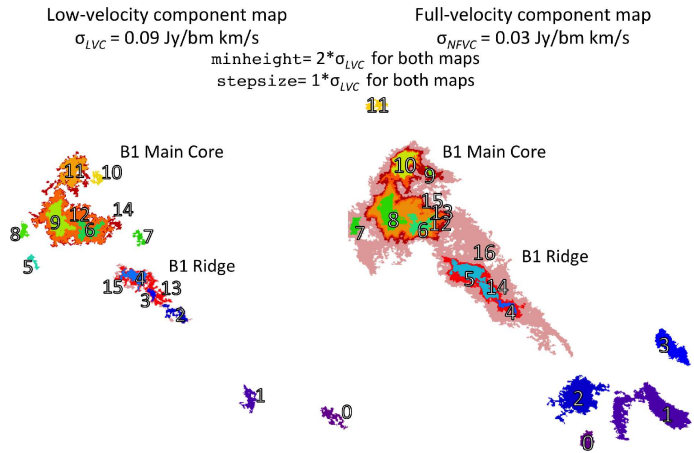}
\caption{\small Structures identified by our non-binary dendrogram
  algorithm from the Barnard~1 integrated intensity maps made from the
  lowest-velocity hyperfine component (left) and from all hyperfine
  components normalized as discussed in the text (right).  The
  dendrogram algorithm parameters are listed in the figure, and are
  the same for each map. Structures are labeled with numbers that
  correspond to leaves and branches in
  Figure~\ref{fig:b1dendrocomparetree2}. See Paper~I for a description
  and analysis of the Barnard~1 region.}
\label{fig:b1dendrocompare2}
\end{figure}

\begin{figure}[!h]
\centering 
\includegraphics[scale=0.095]{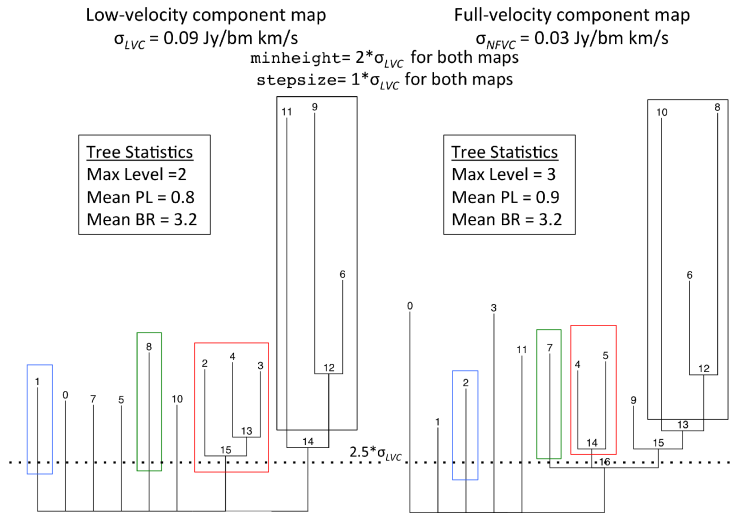}
\caption{\small Non-binary dendrograms from the integrated intensity
  maps made from the lowest-velocity hyperfine component (left) and
  from all hyperfine components normalized as discussed in the text
  (right) using the uniform dendrogram algorithm parameters listed in
  the figure. The colored boxes highlight examples of the same regions
  of emission between the two maps for a comparison of identified
  structures when using uniform algorithm parameters on maps with
  different noise levels. The tree statistics in the figure inset are
  discussed in the text.  See Paper~I for a description and analysis
  of the Barnard~1 region.}
\label{fig:b1dendrocomparetree2}
\end{figure}

For the second case, we used the values of the LVC map for {\tt
  stepsize} and {\tt minheight} for the construction of both
dendrograms. That meant that each dendrogram could branch in
0.09~\Jybm{} steps, and leaves needed to be 0.18~\Jybm{} above their
merge level to be considered real---these were our minimum physically
relevant units to study structure across these two maps with different
noise. The resulting dendrogram-identified structures for the two
Barnard~1 maps are shown in Figure~\ref{fig:b1dendrocompare2}, and the
dendrograms are shown in Figure~\ref{fig:b1dendrocomparetree2}. The
main take-away from these figures is that lowering the noise but
keeping the same absolute units for studying structure results in very
similar dendrograms.  A large-scale, low-intensity branch (structure
16 in the NFVC map) is added around smaller-scale, higher-intensity
structures found in the LVC map, and some leaves that branched
directly from the tree base in the LVC map now branch from one level
up due to the addition of the large-scale, low-intensity branch (e.g.,
leaf 8 becomes leaf 7). The strongest peaks in the region are
similarly identified in both maps (structures 6, 9, and 11 in the LVC
map are 6, 8, 10 in the NFVC map); lowering the noise did not cause
these structures to break up. This provides evidence that if the noise
in a map is lowered, and the {\tt stepsize} and {\tt minheight}
parameters are kept constant, the hierarchical structure will not
significantly change---the main effect is to add lower-intensity
branching structure to connect parts of the emission hierarchy not
previously connected.

Another important test is how the tree statistics change between the
dendrograms when the noise changes. We calculate the tree statistics
of the LVC map in both cases as described in Paper~I, and list the
statistics as insets of Figures~\ref{fig:b1dendrocomparetree1} and
\ref{fig:b1dendrocomparetree2}.  To compare statistics between the LVC
and NFVC maps in each case, we only consider the dendrogram structures
in the NFVC map above the mask level applied to the LVC map.  The
dendrogram of the LVC map has a minimum integrated intensity of
0.21~\Jybmkms{} (set by the 2.5$\sigma_{LVC}$ mask limit to the data
input to the dendrogram algorithm).  Since the NFVC map extends to
lower emission levels set by a 2.5$\sigma_{NFVC}$ mask limit, we
calculate the statistics for that dendrogram above a cut corresponding
to the same 2.5$\sigma_{LVC}$ limit of the LVC map (the cut is
represented in Figures~\ref{fig:b1dendrocomparetree1} and
\ref{fig:b1dendrocomparetree2} by the horizontal dashed line). This
means that leaves from the NFVC map must peak 2$\sigma_{NFVC}$ or
2$\sigma_{LVC}$ above the noisier mask level to be considered real,
for the first and second case, respectively. The leaves that do not
meet this requirement are marked with an ``x'' in
Figures~\ref{fig:b1dendrocomparetree1} and
\ref{fig:b1dendrocomparetree2}. Any branch below this noisier mask
level is removed, and structures above it are lowered one level.

For the first case, where we set {\tt stepsize} and {\tt minheight}
based on the noise-level of the individual map, the NFVC dendrogram
has a maximum branching level three levels higher than the LVC
dendrogram, a mean path length about half a level larger than the LVC
dendrogram, and a mean branching ratio 0.6 higher than the LVC
dendrogram. The differences in maximum branching level and mean path
length can be explained by the NFVC map having a lower absolute
threshold for what can be considered a leaf compared to the
LVC map ({\tt minheight} of 0.06~\Jybmkms{} vs.
0.18~\Jybmkms{}), and allowing branching steps at smaller absolute
units compared to the LVC map ({\tt stepsize} of
0.03~\Jybmkms{} vs. 0.09~\Jybmkms{}); when it is easier to break
emission into leaves and to define new branching levels, there will be
more hierarchical structure in the dendrogram.

For the second case, where we set {\tt stepsize} and {\tt minheight}
based on a minimum relevant \Jybmkms{} value, the mean path lengths
agree to a tenth of a level, the mean branching ratios are identical,
and the maximum branching levels differ by only one level. This
demonstrates how tree structure can be defined so that lowering the
noise of a map will not alter the calculated tree statistics
significantly. When comparing different clouds with different
noise-levels, it will be important to use a minimum relevant value for
algorithm parameters so that the emission in each cloud has the same
absolute threshold for what can be considered a leaf and what can be
considered a significant branching step.

In conclusion, the dendrograms of clouds observed with different noise
can be uniformly compared if the comparison uses absolute units for
the dendrogram contouring and leaf requirements instead of units based
on the noise-level of each individual map, and if the comparison of
tree statistics only analyzes structures above a minimum intensity
level set by the noisiest map.

\end{document}